\documentclass[%
 aip,
 amsmath,amssymb,
 reprint,%
]{revtex4-1}

\usepackage{graphicx}
\usepackage{dcolumn}
\usepackage{bm}

\usepackage[utf8]{inputenc}
\usepackage[T1]{fontenc}
\usepackage{mathptmx}
\usepackage{etoolbox}

\usepackage{xcolor}

\usepackage[normalem]{ulem}

\usepackage{float}

\usepackage{cleveref}
\crefname{figure}{figure}{figures}
\Crefname{figure}{Figure}{Figures}
\crefname{equation}{equation}{equations}
\Crefname{equation}{Equation}{Equations}

\makeatletter
\def\@email#1#2{%
 \endgroup
 \patchcmd{\titleblock@produce}
  {\frontmatter@RRAPformat}
  {\frontmatter@RRAPformat{\produce@RRAP{*#1\href{mailto:#2}{#2}}}\frontmatter@RRAPformat}
  {}{}
}%
\makeatother
\begin{document}

\preprint{AIP/123-QED}

\title{A Useful Metric for the NISQ Era: Qubit Error Probability and Its Role in Zero Noise Extrapolation }
\author{Nahual Sobrino}
 \affiliation{The Abdus Salam International Center for Theoretical Physics (ICTP), Strada Costiera 11, 34151 Trieste, Italy.}
\author{Unai Aseginolaza}%
\affiliation{Basic Sciences Department, Faculty of Engineering, 
Mondragon Unibertsitatea, 20500, Arrasate, Spain.}
\author{ Joaquim Jornet-Somoza}%
\affiliation{Servicios Generales a la Investigaci\'on (SGIker), Universidad del Pa\'is Vasco UPV/EHU, 
            Avenida de Tolosa 72, 
            20018, 
            Donostia/San Sebasti\'an,
            Spain.}
\author{Juan Borge}%
\affiliation{Fisika Aplikatua Saila, Gipuzkoako Ingeniaritza Eskola, University of the Basque Country (UPV/EHU),
Europa Plaza 1, 20018, Donostia/San Sebasti\'an, Spain. }
 \email{juan.borge@ehu.eus}

\date{\today}

\begin{abstract}
Accurate assessment and management of errors is indispensable for extracting useful results from noisy intermediate-scale quantum (NISQ) devices. In this work, we propose the qubit error probability (QEP), a device specific metric that combines relaxation, dephasing, gate, and measurement contributions into a single per qubit figure of merit computable before execution. Leveraging QEP as the control variable, we revisit zero noise extrapolation (ZNE) by adding pairs of controlled native two-qubit gates on all connected qubit pairs to generate circuits with successively larger mean QEP; the zero error limit is then approximated by a linear regression of the measured observable against those values.

Benchmarking on IBM Quantum Heron processors, we apply QEP guided ZNE to first order Trotterized simulations of the two dimensional transverse field Ising model, chosen as a representative interacting many body system, involving up to 68 qubits and 15 Trotter steps. In regimes where the raw circuits exhibit a finite mean QEP, the method suppresses observable errors beyond those attainable with circuit depth scaled ZNE, while requiring only three noise scaled evaluations and no additional classical post processing. These results demonstrate that QEP serves as a transparent and efficient error metric, and that its integration into ZNE provides a practical route to reliability gains on current superconducting hardware, without the resource costs associated with full quantum error correction.
\end{abstract}

\maketitle

\section{Introduction \\}

Quantum computing, defined as the ability to harness real quantum states to perform complex calculations, has recently transitioned from a theoretical possibility to a practical reality \cite{Nielsen2011}. Feynman’s proposition \cite{feynman1982} of using quantum systems to simulate quantum mechanics is no longer a conceptual vision but a tangible achievement. Over the past two decades, the capabilities of advanced quantum computers have seen remarkable progress. In principle, ideal quantum computers are expected to execute calculations beyond the reach of classical computers with exceptional precision. However, achieving this accuracy requires devices with thousands of qubits to enable effective quantum error correction \cite{lidar_brun_2013,TerhalRMP2015,Wendin_2017}. Unfortunately, this goal remains distant. In the current noisy intermediate-scale quantum (NISQ) era, scientific and technological efforts are focused on evaluating, controlling, and mitigating physical errors \cite{BhartiRMP2022,Leymann_2020,Porter2022,Kandala2017,Aspuru-Guzik-Mol-Science2005,Cerezo2021,Arute2019,Preskill2018,Xiao2021,Dalzell2020,georgopoulos2021modeling,patel2020experimental,nation2021scalable,Weidenfeller2022,SetiawanPRX2021,WuPRL2021,HeadleyPRA2022}.

NISQ devices typically consist of a few dozen qubits and are expected to scale to a few hundred in the near future. A defining characteristic of these devices, as their name implies, is their inherent noisiness. Currently, some of the most advanced quantum computers are based on transmon qubits \cite{transmon,natureIBM2023,natureGoogle2024}. However, these systems face five significant limitations. First, the stability of a qubit state in these devices is typically of the order of hundreds of microseconds, thus restricting the duration of feasible computations. Second, the operations (gates) that act on the qubits are noisy, leading to a progressive loss of accuracy during calculations. Third, the process of measuring the qubit state is prone to errors. Fourth, there is a possibility that an operation on one qubit affects another, the crosstalk. Finally, the number of available physical qubits remains limited, imposing further constraints on computational capability.

The error estimation in a quantum calculation, usually requires comparing the result with the corresponding from a classical calculation. Nevertheless, quantum computers in the NISQ era are able to perform state-of-the-art classical computations \cite{natureIBM2023,natureGoogle2024} and it is expected that they may overcome them in the near future.  For this reason, it is preferable to avoid relying on classical computations to determine the error in quantum calculations. Therefore, we need to find metrics to measure errors that avoid the use of classical simulations. Sometimes we can use specific Clifford points \cite{natureIBM2023} where the exact analytical result is known and then extrapolate the results to the non-trivial points. However, there are many situations in which it is not possible to compare this with any analytical result. In a previous work \cite{Aseginolaza2024} we proposed a pre-processing tool (the Tool for Error Description in Quantum Circuits, TED-qc), which computes the probability of an error to occur in a quantum circuit provided the hardware and its calibration, without any actual evaluation of the circuit. This tool has proven to be a good estimator of the error of a quantum circuit \cite{brody2025grover,gardeazabal2024machine,antero2025robot,de2025fidelity}.  However, when a calculation involves many qubits, its usefulness may be limited, as errors in the quantum circuit do not necessarily impact the final result in all cases. In this paper, we introduce a more precise metric for assessing the actual error in a quantum computation. Rather than considering only the total error probability, we focus on the individual qubit error probability (QEP), which provides a more refined measure of the error’s impact on the calculation.

As noted above, quantum computing faces significant challenges due to noise and errors that arise from imperfections in hardware, environmental interactions, and quantum decoherence. Unlike classical computers, which can use traditional error correction methods, quantum systems require specialized techniques to mitigate noise while preserving quantum coherence. Error mitigation techniques aim to improve the accuracy of quantum computations in NISQ devices at the cost of increasing the quantum computation time. This may allow researchers and industry professionals to extract meaningful results despite the presence of noise. These techniques do not eliminate noise entirely, but instead reduce its effects, making quantum algorithms more reliable and practical for real-world applications.

Several error mitigation strategies \cite{cai2023quantum} have been developed to address different sources of error, including Probabilistic Error Cancellation (PEC) \cite{natureIBM2023}, which leverages classical post-processing to counteract noise by applying carefully designed inverse transformations, and Dynamical Decoupling, which use precisely timed control pulses to suppress decoherence and extend qubit coherence times. Additionally, measurement error mitigation techniques correct readout errors by characterizing and inverting the noise model affecting quantum measurements.

Among the error mitigation techniques, one of the most widely used is the Zero-Noise Extrapolation (ZNE) \cite{li2017efficient,Mitigation-PRL-2017,giurgica2020digital}, which systematically amplifies noise in a controlled manner, executes the quantum circuit under varying noise regimes, and extrapolates the computed results to approximate an idealized zero-noise limit. This approach relies on the assumption that the quantum system's response to noise follows a predictable mathematical trend, enabling the reconstruction of an estimated noiseless expectation value. The fundamental principle of ZNE is to employ numerical extrapolation methods —such as linear, polynomial, or exponential regression— to model the dependency of the circuit's output on noise strength and infer the limit where noise approaches zero. By artificially scaling noise levels, typically through pulse stretching, gate repetition, or hardware-level noise injection, ZNE enables the extraction of noise-free estimates without requiring full quantum error correction. This technique has demonstrated effectiveness in improving the fidelity of quantum computations in domains such as quantum chemistry simulations, variational quantum eigensolvers, and other near-term quantum algorithms where mitigating errors is essential \cite{natureIBM2023,uvarov2024mitigating,garmon2020benchmarking,keen2020quantum,mitigation_npj_2021,ZNE_PRA2020}. Importantly, while the computational cost of techniques such as PEC or Richardson extrapolation \cite{Richardson_extr_PRA2022, Richardson_extr_PRA2024} increases exponentially with the number of qubits, ZNE remains independent of qubit count, typically requiring only a error amplification factor of 3-5 in additional quantum computational resources. This scalability makes it one of the most applied quantum error mitigation techniques.

The key issue regarding ZNE is how errors are quantified and amplified. The usual procedure, like the one implemented in frameworks such as Qiskit, is to consider that the original circuit has an error of one. Then, copies of the circuit are introduced in pairs, preserving the computed result while increasing the overall error. These methods use a metric that considers the error in a simplistic manner; for instance, if a new circuit is built with three copies of the original circuit, it is assumed that the error is three times the original. This is a dubious approximation since the error of a circuit does not increase linearly with the circuit depth as we will see later on. In this paper, we propose to use the mean qubit error probability (QEP) as a metric to quantify and control error amplification. This metric represents more accurately the error that occurs in quantum circuits. 

Our benchmark circuit to test both the QEP and its effect in ZNE is the Trotterized time evolution of a two-dimensional transverse-field Ising model. The Ising model is widely studied across various domains of physics and has been extended in innovative ways in recent simulations investigating quantum many-body phenomena, including time crystals, quantum scars, and Majorana edge modes\cite{CerveraIsing2018,mi2022time,frey2022realization,chen2022error}.

Specifically, we examine the time evolution of the Hamiltonian:
\begin{equation}
\label{hamiltonian}
    H=-J\sum_{\left< i,j\right> } Z_iZ_j + h\sum_i X_i,
\end{equation}
where $J>0$ is the coupling of nearest-neighbor spins with $i<j$ and $h$ is the global transverse field. Here, $Z_i$ and $X_i$ represent the Pauli-$Z$ and Pauli-$X$ operators acting on qubit $i$, respectively. The time evolution of an initial quantum state can be approximated using a first-order Trotter decomposition of the time-evolution operator, which factorizes the Hamiltonian dynamics into discrete steps, in the following way:
\begin{eqnarray}
e^{-iH_{ZZ}\delta_t} &=& \prod_{\left< i,j\right> }e^{iJZ_iZ_j\delta_t} = \prod_{\left< i,j\right> }R_{Z_iZ_j}(-2J\delta_t ) \nonumber \\
e^{-iH_{X}\delta_t} &=& \prod_{i}e^{-ihX_i\delta_t} = \prod_{i}R_{X_i}(2h\delta_t).
\label{trotter}
\end{eqnarray}
where $H_{ZZ} = -J\sum_{\left< i,j\right> } Z_iZ_j$ and $H_X = h\sum_i X_i$ denote the two separate terms of the Hamiltonian, which are independently exponentiated in the Trotter decomposition. The evolution time $T$ is discretized into $T/\delta t$ Trotter steps, and  $R_{Z_iZ_j}(\theta_j)$ and $R_{X_i}(\theta_h)$ are the ZZ and X rotation gates, respectively.  Each Trotter step sequentially applies the two evolution operators defined above, and the total circuit depth increases proportionally with the number of Trotter steps. This flexibility allows for controlled simulation accuracy by increasing the number of steps.

Following the lines of Kim Y. et al.\cite{natureIBM2023} we study the the total magnetization $M = \sum_i\langle Z_i \rangle$ for two different setup configurations. In the first one, we fix $\delta_t=1/4$, $J=\pi$, and $h=0$ to position the system at a Clifford point, allowing for an efficient comparison between our simulations and the exact analytical solution. In this case, we perform calculations up to 68 qubits and vary the number of Trotter steps. In the second one, we fixed the number of Trotter steps to $15$, $\delta_t=1/4$, $J=\pi$ for different values of $h$. As we are not at a Clifford point we have done the calculation up to $32$ qubits in order to be able to simulate the theoretical values in a classical computer. 

The paper is organized as follows. In \cref{ted-qc}, we introduce the qubit error probability (QEP), and compare it with other metrics. In \cref{ZNE}, we use this QEP to track the error in order to improve standard ZNE. In \cref{results}, we compare the results obtained using our new tool against prior methods. Finally, in \cref{Conlusions}, we summarize our conclusions.

\section{\label{ted-qc} Qubit error probability}

As discussed in the previous section, error estimation is crucial in the NISQ era of quantum computing. In Aseginolaza U. et al.\cite{Aseginolaza2024}, we calculated the total error probability and, following the same scheme, we will calculate the qubit error probability (QEP). As we know, errors arise primarily from four sources:

\begin{itemize}
\item The error induced by the instability of the qubits. In order to have non-trivial states we have to excite our physical qubits. The probability of finding the qubit in the excited state decays exponentially with time \cite{rigetti2012superconducting,place2021new}, so the larger the quantum circuit the less likely is to find the qubit in the expected state. There are two possible decaying mechanisms, the decay of an excited state to the ground state, i.e. the probability of a state $|1\rangle$ to decay to a $|0\rangle$; and the change of the phase of an excited
state, for example passing from the state $1/\sqrt{2}(|0\rangle+|1\rangle)$ to the state $1/\sqrt{2}(|0\rangle-|1\rangle)$.
\item Gate errors: each quantum gate introduces errors with single-qubit gates typically having error probabilities of the order of $10^{-4}$ to $10^{-3}$ and two-qubit gates of the order of $10^{-3}$ to $10^{-2}$.
\item Measurement errors: the act of measuring qubits introduces additional inaccuracies, typically on the order of $10^{-2}$ to $10^{-1}$.
\item The crosstalk between different qubits: when we act on a qubit we may act unintentionally on another. This type of error is very difficult to track, but luckily the latest IBM processor, \textit{Heron}, have almost zero crosstalk \cite{sheldon2016procedure,dial2022eagle,chow2021ibm}. All the results shown in this paper have been calculated on IBM Quantum Heron processor on the $ibm\_torino$ quantum system.
\end{itemize}

In order to calculate the qubit error probability we define the success probability of the j-th qubit, i.e. the possibility of non-committing any error on the j-th qubit, $S_{j}$, as
\begin{align}
	\label{St}
	S_j=\prod_{i=1}^{m}S_{ij}=\prod_{i=1}^{m} [1-P_{ij}],    
\end{align}
where $P_{ij}$ represents the error probability from any source acting on the j-th qubit: single and double qubit gates, measurement, and qubit instabilities; where $m$ is the total number of error sources. The relation between error probability and success probability is defined as $P+S=1$. The qubit error probability, QEP is therefore computed as
\begin{equation}
	P_j= 1- [1- P_{j}^{\mathrm{meas}}][1-P_j^{\tau_1}][1-P_j^{\tau_2}]\prod_{i=1}^{\mathrm{gates}}[1-P_{ij}^{\mathrm{gate}}].
	\label{eq_S}
\end{equation}
The first term is the measurement error for the j-th qubit, denoted as $P_{j}^{\mathrm{meas}}$, and the second and third terms are the errors arising from qubit instability, represented by $P_j^{\tau_1}$ and $P_j^{\tau_2}$. Finally, the error due to all gates applied/affecting on the j-th qubit are considered on the last term, $P_{ij}^{\mathrm{gate}}$. While the gate and measurement errors are directly obtained through the IBM's API, to account for errors caused by instability qubits $P_j^{\tau_i}$ we consider an exponential decay of the form \cite{abraham2019qiskit}
\begin{equation}
	P_j^{\tau_i}=1-e^{-\frac{t_{j}}{\tau_{i,j}}}
    \label{eq_P_tau}
\end{equation}
where  $i=1,2$ stand for the two instability mechanism, relaxation and dephasing explained before, $\tau_{i,j}$ are the related times for the j-th qubit, and $t_{j}$ is the total circuit time of j-th qubit. The evaluation of \cref{eq_P_tau} requires the knowledge of the time it takes for any qubit between the initialization of the circuit and the measurement $t_{j}$. This is done by adding up how long it takes for the j-th qubit  to perform any single gate.
The situation becomes more complex when dealing with two-qubit gates. The process of executing a two-qubit gate necessitates the completion of all preceding gate operations on both qubits before the joint operation can be executed. Consequently, after incorporating the time needed for the two-qubit gate into the total time calculation, we are faced with the task of comparing the cumulative times of the two qubits involved.
The crucial point here is that we don't simply add these times together. Instead, we analyze the total times for each qubit and then adopt the longer of the two as the updated qubit time. This approach ensures that we are accounting for the maximum time it could take, thus capturing the full potential for instability-induced error.

The last product in \cref{eq_S} accounts for all gate-related errors $P_{ij}^{\mathrm{gate}}$ (both single and double) that affect the j-th qubit. There are two different gates that may produce an error on the j-th qubit. On the one hand, we have the gates acting directly on it. On the other hand, we have the possibility that the control qubit on a two-qubit gate suffered an error while acting on it. In order to take this into account, we add the gates affecting the control qubit gate to the ones affecting the j-th one.  

These errors depend on the specific quantum hardware, qubits and connections used. To address this, we developed TED-qc (Tool for Error Description in Quantum Circuits), an open-source pre-processing tool available on GitLab  (\href{https://gitlab.com/qjornet/ted-qc/}{https://gitlab.com/qjornet/ted-qc/}). Currenlty written for Qiskit v1.2.0 language, it analyzes quantum circuits, extracts error probabilities from IBM’s API and computes total error probabilities. The code can be easily extended to any quantum computer using its calibration parameters. 

\begin{figure*}
  \centering  
  \includegraphics[width=0.8\textwidth]{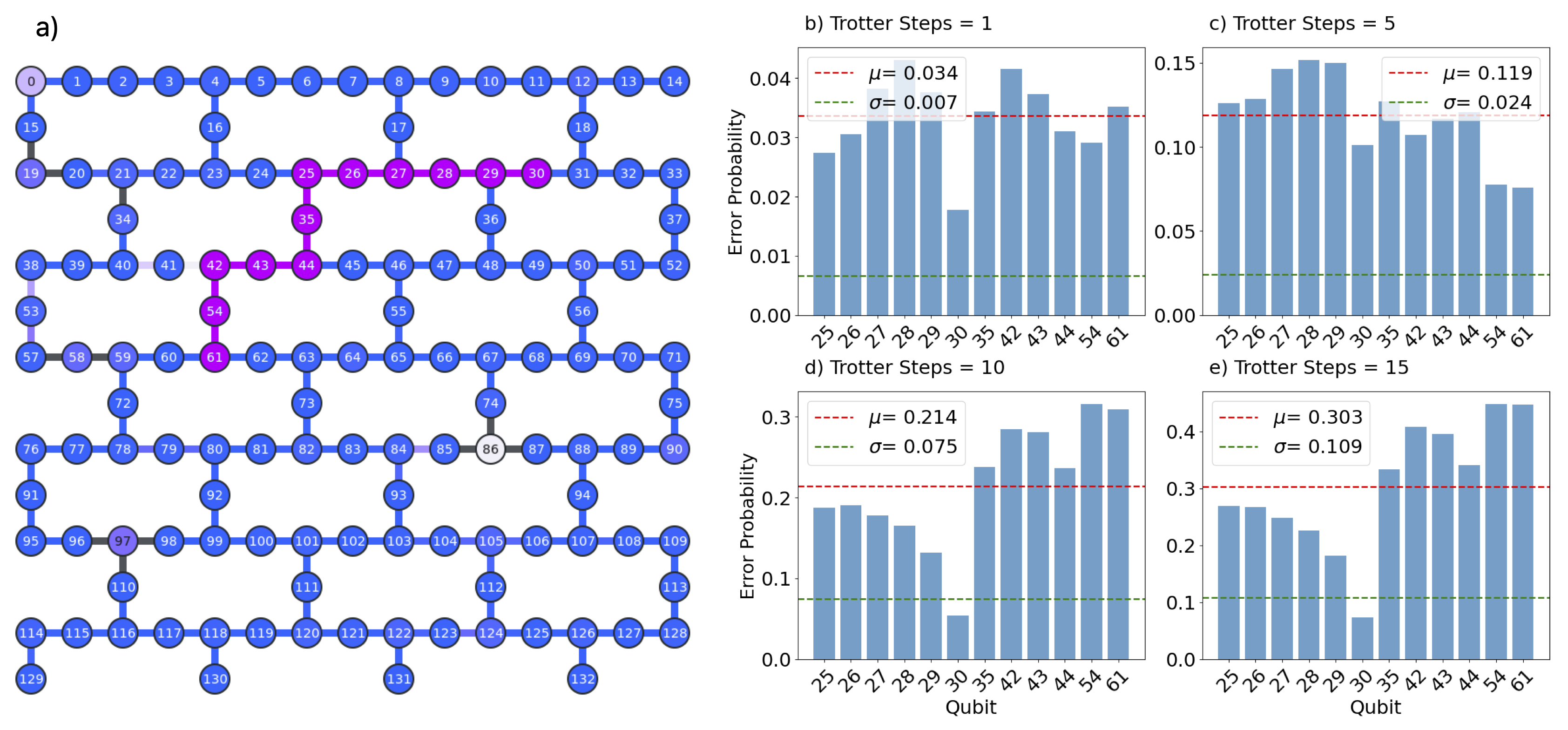}
  \caption{Schematic representation of the procedure to get the mean Qubit Error Probability. a) 12 phyiscal qubit mapping on $ibm\_torino$ quantum system for the 12 site Ising Hamiltonian (\cref{hamiltonian}). b) to e) Computed Qubit error probability (QEP) of each qubit in a circuit with $1$, $5$, $10$, and $15$ Trotter steps, respectively. Dashed red and green lines represent the mean ($\mu$) and standard deviation ($\sigma$) of the QEP values, respectively.}
  \label{Prob-error}
\end{figure*}

The calibration procedure of a quantum hardware can detect failed qubits and gates, and sometimes can return inaccurate values. We have included a warning in the code to be aware of the possibility of a gate being poorly calibrated. When a two-qubit gate has an error that is $100\%$ larger than the mean value of all the two-qubit gates involved in the circuit, we highlight this issue as poorly calibrated, since it can induce a huge qubit error probability. Sometimes there is no information regarding a particular gate, for example between the $85$th and $86$th qubits in \cref{Prob-error}a, hence QEP gives an error probability of $1$, which obviously may be misleading.

\Cref{Prob-error} shows an example of how QEP values are obtained in our tool. We have run a $12$-qubit trotterization of Ising Hamiltonian expressed in \cref{hamiltonian} with $\delta_t=1/4$, $J=\pi$, and $h=0$ for different number of Trotter steps on $ibm\_torino$. In the left panel a), we can see the 12-qubit chain chosen by the Qiskit \texttt{transpile} function required to map the logical qubits into the physical quantum hardware. In the right panels, b) to e), we can see the error probability of each qubit in the circuit for $1$, $5$, $10$, and $15$ Trotter steps, respectively. We can see how the mean ($\mu$) and variance ($\sigma$) of the QEP increase as the number of Trotter steps are increased. It is also clear from the image that different qubits have different QEP; however, since the variance is one order of magnitude smaller in all the cases, we take the average value as the extrapolation variable for the error mitigation strategy.

\section{\label{ZNE} Using Qubit Error Probability in the Zero Noise Extrapolation}

\begin{figure*}
  \centering  
   \includegraphics[width=0.8\textwidth]{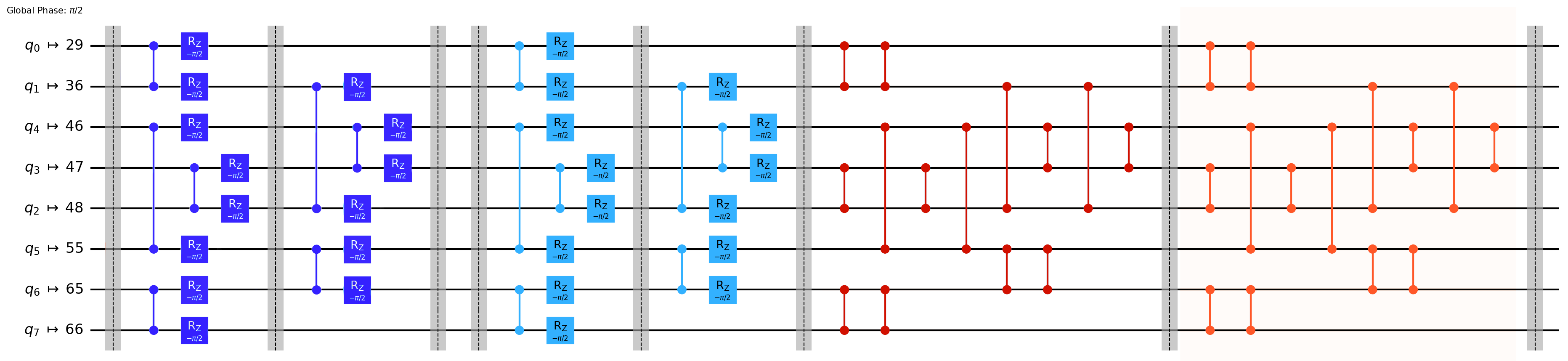}
  \caption{Transpiled quantum circuit for the Ising Hamiltonian evolution (\cref{hamiltonian}) for a 8 qubits system in the Clifford point ($J=\pi$, and $h=0$). In blue are represented the basic gate operations used in $ibm\_torino$ quantum system for two (dark and light blue) trotterization steps.   In red, the CZ entangle gates used for the error amplification are represented. The different color intensity accounts for each of the two error amplification factor.} 
  \label{ZEPE_circuit}
\end{figure*}

Once we have defined the QEP, we can use it as the extrapolation variable in the ZNE method.  The idea is the same as in standard ZNE: we execute multiple circuits with the same expectation value as the original one but ensuring that the replicas will have larger error probabilities each time. To do so, we insert pairs of controlled two-qubit native gates into all connected qubits in the circuit, which in the case of $ibm\_torino$ are controlled-Z gates (CZ). We know that a pair of these two-qubit gate is just a unitary operation, so the expectation value remains the same as the original one. The addition of this set of gates defines a factor of one. We can add as many factors as we want to further amplify the noise, and hence the error probability (see \cref{ZEPE_circuit}). As we pointed out before, we use the mean QEP to measure the "error", and then we extrapolate the mean QEP to zero to correct our result.

It is important to note that, as in the standard ZNE, these calculations are compatible with other error mitigation techniques. Here, we have used the T-REX mitigation technique \cite{van2022model,chen2019detector,bravyi2021mitigating} as implemented in Qiskit to eliminate measurement errors. This is particularly relevant because when we use ZNE we cannot increase the error due to the measurement because the final state would not be invariant, meaning it cannot be mitigated within this framework. Since we have mitigated this source of error by applying T-REX, we did not take it into account while calculating the QEP, assuming it has been largely reduced. 

ZNE has been shown to perform better when combined with twirling techniques \cite{Twirling2016}.  Since we aim to compare our method against standard ZNE, we have also incorporated this error mitigation technique in our calculations. Another important issue to take into account is the possible calibration problems explained above. Then, the role of warnings identifying if a qubit suffers from calibration issues or unexpected behavior is crucial, since the QEP may be miscalculated, leading to ineffective mitigation.
\begin{figure}
  \centering  
  \includegraphics[width=1.\linewidth]{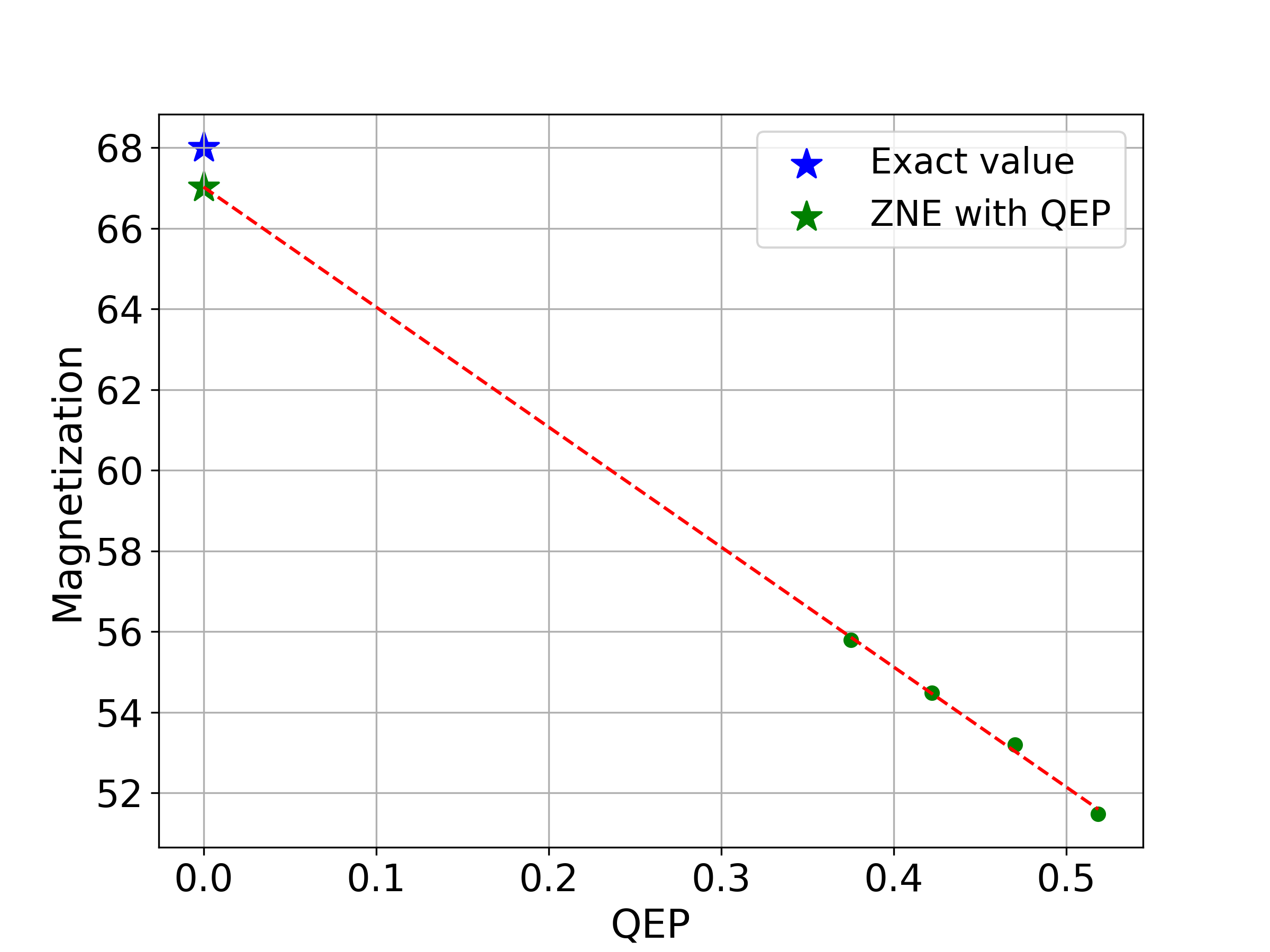}
  \caption{Use of QEP to mitigate the error  in calculating the magnetization of a 68 qubit system after $11$ Trotter steps with $\delta_t=1/4$, $J=\pi$, and $h=0$ in $ibm\_torino$. The green points are the raw, factor $1$, factor $2$, factor $3$ values of the magnetization as a function of the mean qubit  error probability (QEP). The blue star represents the theoretical exact result, and the dashed red line is the linear error extrapolation. Finally, the green star stands for the mitigated expectation value obtained from the extrapolation to the zero-error probability.}
  \label{ZEPE}
\end{figure}

In \cref{ZEPE}, we calculate the magnetization of an $11$-Trotter-step Ising system as a function of the average QEP. The first green point represents the result of the circuit without any ZNE  mitigation techniques. The other three green points correspond to the results obtained after adding a set of two-qubit gates as indicated above (factor $1$, factor $2$, and factor $3$). The blue star is the theoretical exact value, and the dashed red line represents the zero error probability extrapolation. We can see that this method effectively mitigates noise mostly due to the gate errors and qubit state decay. It is important to stress that this method, like standard ZNE, is heuristic, so we cannot  guarantee that it will be able to reduce all the noise.
\begin{figure}
  \centering  
  \includegraphics[width=1.\linewidth]{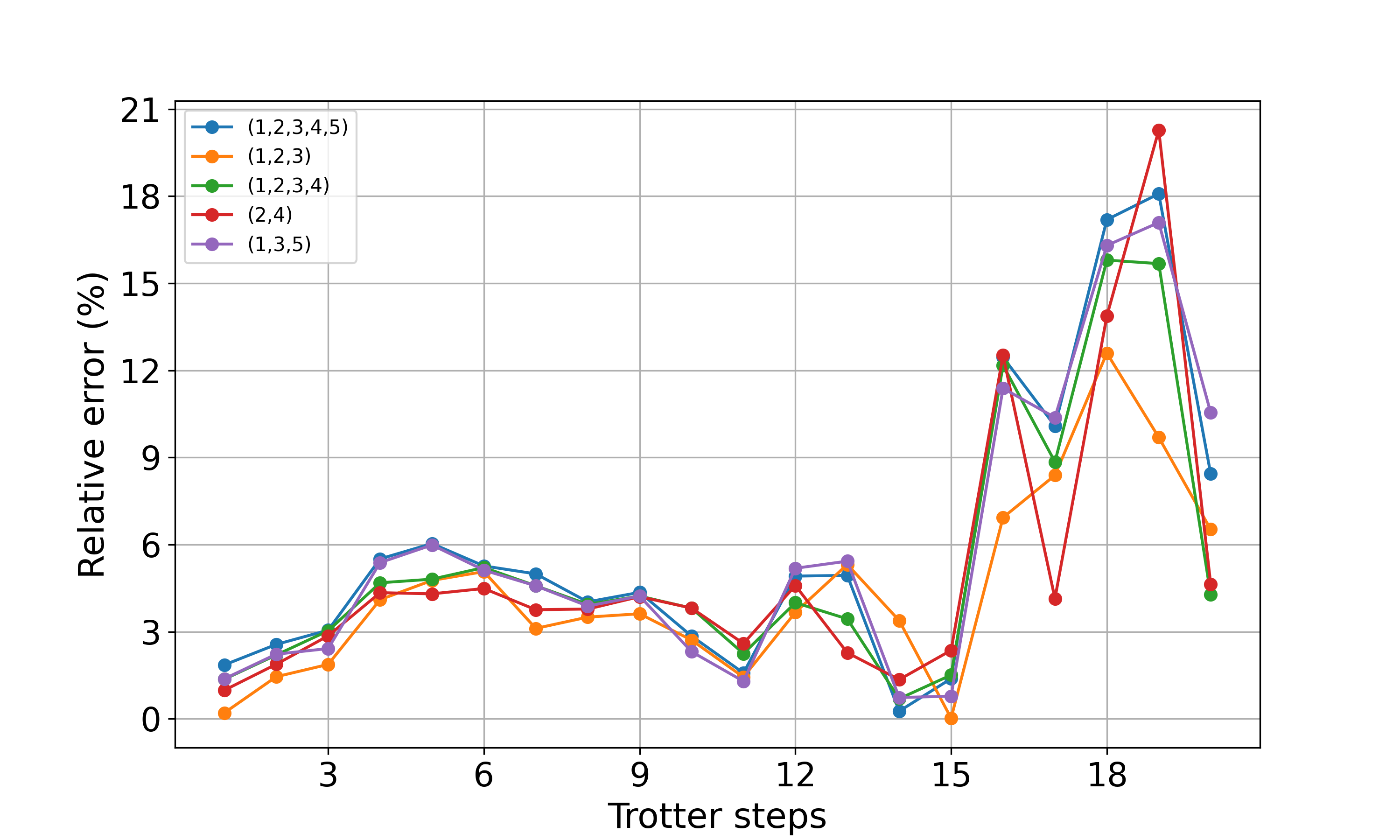}
  \caption{ZNE with QEP magnetizations with $\delta_t=1/4$, $J=\pi$, and $h=0$ as a function of the number of Trotter steps. Different colors correspond to different extrapolation adjustments, i.e. using different set of error amplification factors. For instance, blue line labeled as (1,2,3,4,5) corresponds to the linear extrapolation performed with the raw, factor 1, factor 2, factor 3, factor 4, and factor 5 points.}
  \label{factor}
\end{figure}

Finally, we have analyzed ZNE as a function of the error amplification factor. Different factors can be used to perform the extrapolation. We have run several quantum circuits increasing the number of Trotter steps using amplification error factor ranging from 1 to 5. Then, we used different 
combination of the expectation values of the amplified circuits to do the error zero probability extrapolation. \Cref{factor} shows the results of the extrapolated values using different error amplification factor for the QEP calculation. We can see  that the best mitigation corresponds to the use of "raw" and $1$, $2$, and $3$ factors to extrapolate. 

Among all the extrapolation options, we have chosen to use the linear extrapolation for two reasons.  First, we may expect that the error while calculating most operators, especially the magnetization, scales linearly with the mean QEP. Second, we aim to avoid extrapolations such as the exponential one, which can induce to a significant deviation from the actual expectation value. This is due to the fact that an small error calculating the exponent may induce a huge error in the value we are extrapolating.

\section{Comparison between standard ZNE and ZNE with QEP}
\label{results}
 Once we have developed the ZNE with QEP calculation method, we aim to compare it with the standard ZNE function implemented in Qiskit. To do so, we will compute the previously explained trotterization for the Ising Hamiltonian for two different cases. In the first one, we compute a 68-qubit chain in $ibm\_torino$ quantum system for different numbers of Trotter steps fixing $\delta_t=1/4$, $J=\pi$, and $h=0$. As we pointed before this correspond to a Clifford point so we can easily know the exact value. The second case we will fix $\delta_t=1/4$, $J=\pi$ and the number of Trotter steps to $15$, changing the values of $h$. This corresponds to non-Clifford points so we will do it for a $32$ qubit system in order to be able to simulate it classically. Using QEP has two main advantages in comparison with standard ZNE. First, we have enhanced control over error and noise, utilizing the quantum computer's calibration to more precisely monitor noise behavior. Second, increasing the noise is more straightforward and affordable, as it can be achieved by adjusting the circuit depth which enable us to apply it for more complex circuits.

\begin{figure}
  \centering  
  \includegraphics[width=1.\linewidth]{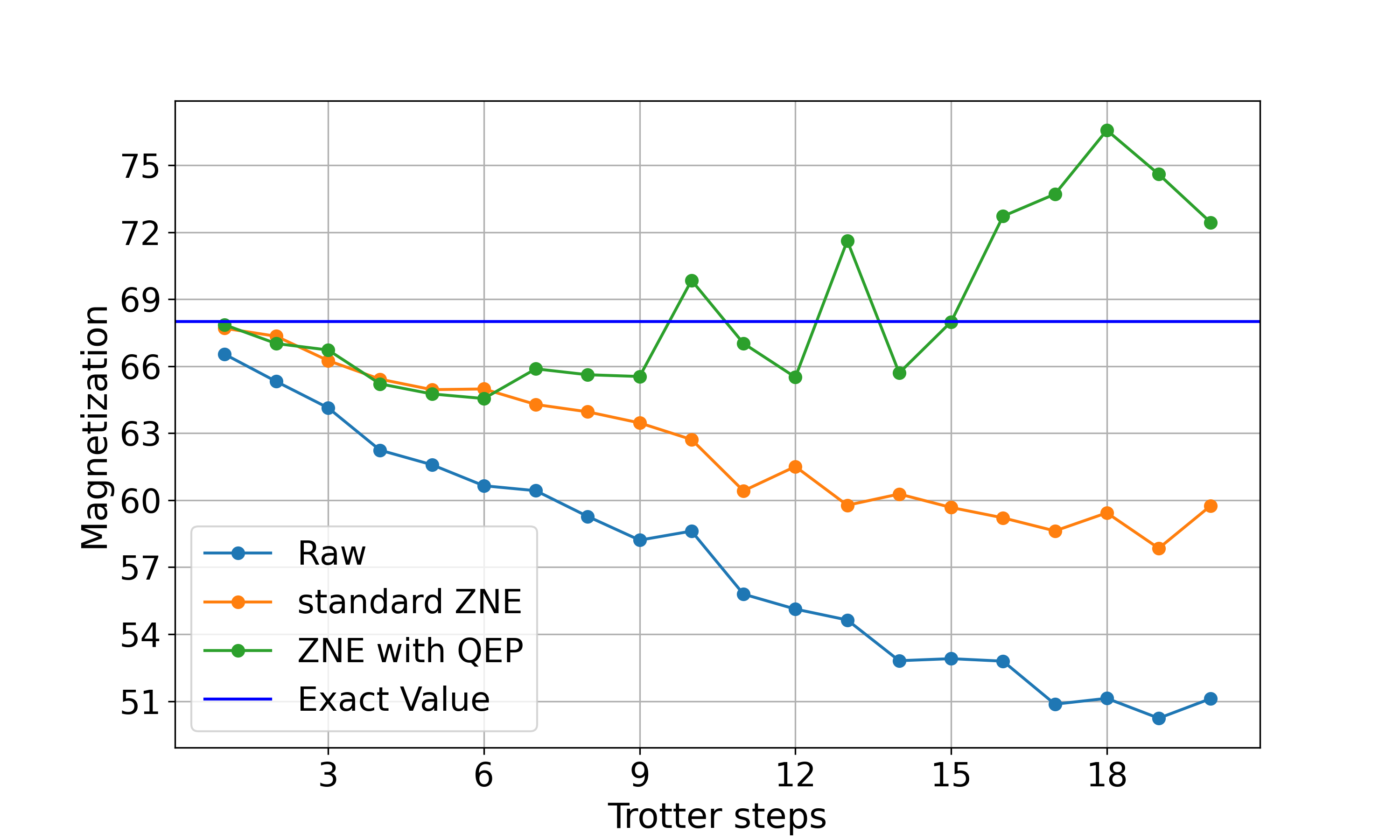}
  \includegraphics[width=1.\linewidth]{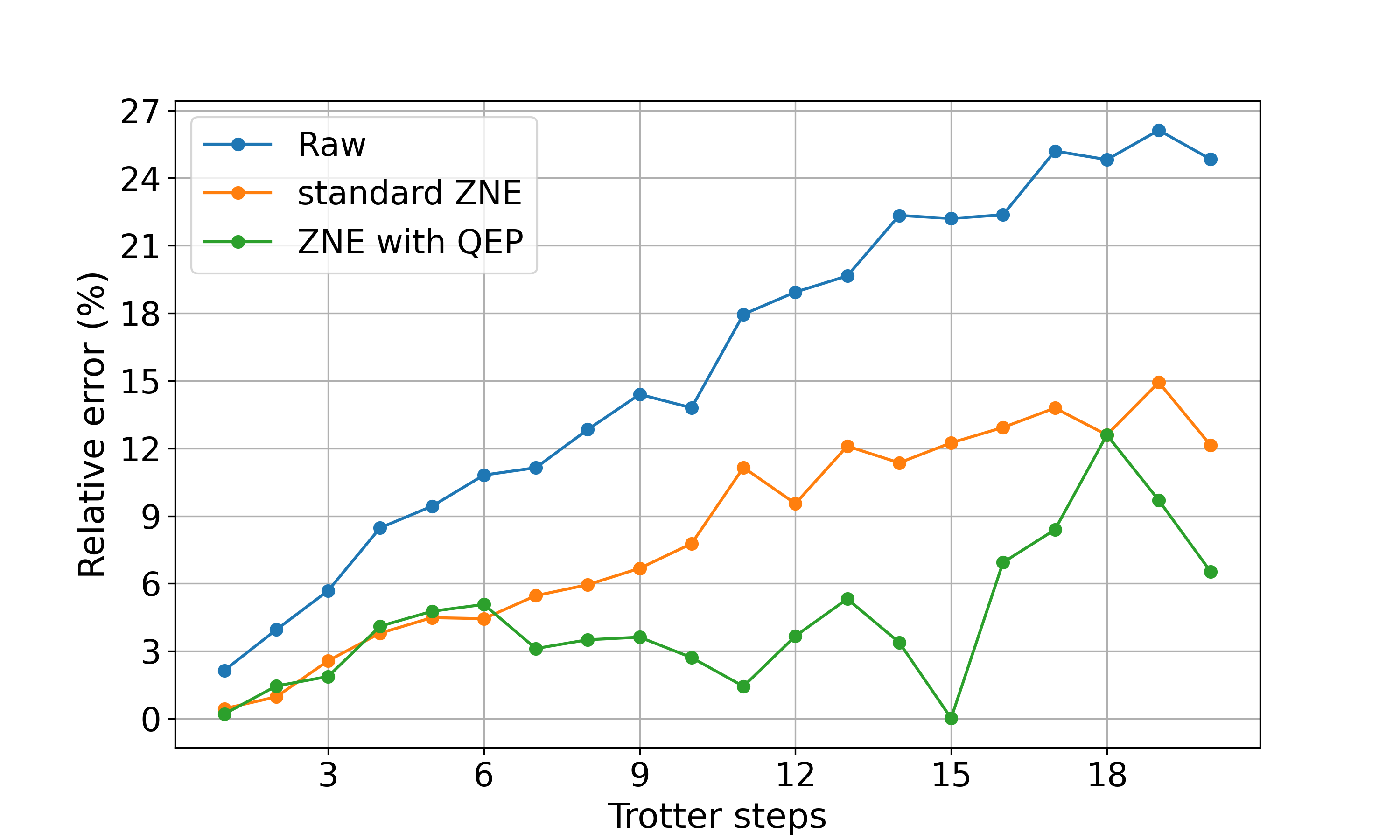}
  \caption{Comparison on the reliability between the different error mitigation techniques. Left: Absolute value of the total magnetization as a function of the amount of Trotter steps. The raw (orange), standard ZNE (blue), ZNE with QEP (green) and exact value (solid line at $M=68$) are included. Right: Relative error of the raw and extrapolated values compaired with the exact theoretical result.}
  \label{absolute}
\end{figure}

\begin{figure}
  \centering  
  \includegraphics[width=1.\linewidth]{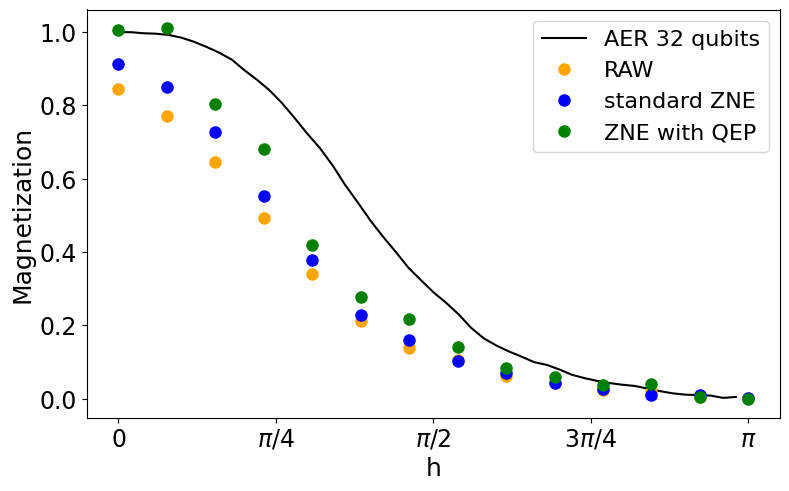}
  \caption{Comparison on the reliability between the different error mitigation techniques.  Absolute value of the total magnetization as a function of the $h$ parameter. The raw (orange), standard ZNE (blue), ZNE with QEP (green) and exact value (solid line) are included.}
  \label{fig_h}
\end{figure}

In \cref{absolute}, we can see the absolute magnetization (up) of the raw, i.e. without ZNE correction, standard ZNE, and ZNE with QEP results, and the absolute error (bottom) as a function of the number of Trotter steps. We can see that our method reduces the error more effectively than standard ZNE. It performs better for $6$-$16$  Trotter steps. The effectiveness of the method directly depends on the entangle depth of the original quantum circuit. 
On the one hand, if the raw circuit has a small error probability, i.e. short depth, we cannot mitigate a significant portion of it, since the currently NISQ quantum machine is accurate enough to handle it. On the other hand, if the raw circuit has a mean QEP larger than $0.6$, the error due to circuit time and the error propagation on the gate operations starts to be sufficiently large for an effective mitigation. In this regime, ZNE with QEP performs better than standard ZNE because the latter suffers of incommensurate circuit time execution which yields in an inaccurate manner to account for noise.  

In \cref{fig_h}, we present the absolute magnetization for the raw, standard ZNE, and ZNE with QEP results, as a function of the $h$ parameter. As in the previous case, the inclusion of QPE improves the effectiveness of standard ZNE. 

We have shown that the use of QPE in ZNE can be an easily tuneable error mitigation technique. It requires minimal quantum computational resources and performs better than standard ZNE. This new method has been used on IBM quantum computers but can be extended to any type of hardware.

\section{Conclusions}
\label{Conlusions}

In this work, we have investigated the primary sources of errors in quantum computers to define the QEP, a fundamental metric for quantifying the likelihood of qubit errors. Given the inherent noise and instability in NISQ devices, to accurately estimate and mitigate these errors is essential for improving quantum computation reliability. The QEP metric provides a systematic approach to assessing errors at the qubit level, offering valuable insights into error propagation and its impact on quantum circuits. By leveraging QEP, we aim to enhance the precision and effectiveness of quantifying the  error  and mitigate it in quantum computing.

One of the most critical aspects of the NISQ era is the development of efficient error mitigation techniques to bridge the gap between noisy quantum hardware and the desired quantum advantage. In this context, we explored how QEP can be utilized to improve one of the most commonly used techniques, the zero-noise extraploation ZNE. Standard ZNE methods attempt to infer error-free results by executing quantum circuits at different noise levels and extrapolating back to a noiseless regime. However, these approaches have limitations, particularly in their sensitivity to noise models and the complexity of implementation. By integrating QEP into this process, we improve the standard ZNE approach and enhance its overall performance.

The results of our study demonstrate that the inclusion of QPE offers significant advantages over traditional ZNE methods. As standard ZNE, and in contrast with other mitigation techniques, it requires minimal quantum computational resources while achieving higher accuracy in error mitigation. Additionally, it is easily tunable, allowing users to optimize its parameters based on the specific noise characteristics of a given quantum system. These attributes make it particularly valuable for practical implementations where computational overhead must be minimized while maintaining high levels of precision.

Although our research has primarily focused on IBM quantum computers, the method is not hardware-specific and can be applied to other quantum computing platforms. This flexibility ensures that it can contribute to a wide range of quantum technologies,  controlling their errors and enhancing their error resilience and overall performance through the improvement of ZNE. Future work could explore further refinements to the technique, including its integration with other error mitigation methods and its application to more complex quantum algorithms. By advancing robust and efficient error mitigation strategies,  we move closer to achieving practical quantum utility, where quantum computations can rival and surpass classical methods in real-world applications.

\section*{Acknowledgments}
We acknowledge support from the Basq initiative. 
This work has been possible thanks to the quantum computational resources provided by the BasQ Strategy under the collaboration agreement between Ikerbasque Foundation and the University of the Basque Country (UPV/EHU) and the University on Mondragon, on behalf of the Department of Science, Universities and Innovation of the Basque Government. We acknowledge the use of IBM Quantum services for this work. The views expressed
are those of the authors and do not reﬂect the ofﬁcial policy or position of IBM or the IBM Quantum team.
We also acknowledge support from the Basque Goverment ELKARTEK KUBIT project with reference KK-2024/00105. N.S. acknowledges funding from the European Union under the Horizon Europe research and innovation programme (Marie Skłodowska-Curie grant agreement no. 101148213, EATTS).

\section*{Author Declarations}
\subsection*{Conflict of Interest}
The authors have no conflicts to disclose.

\section*{Data Availability}
The data that support the findings of this study are available from the corresponding author
upon reasonable request.

\section*{References}
\nocite{*}
\bibliography{refs_SJ-3}

\begin{thebibliography}{79}%
\makeatletter
\providecommand \@ifxundefined [1]{%
 \@ifx{#1\undefined}
}%
\providecommand \@ifnum [1]{%
 \ifnum #1\expandafter \@firstoftwo
 \else \expandafter \@secondoftwo
 \fi
}%
\providecommand \@ifx [1]{%
 \ifx #1\expandafter \@firstoftwo
 \else \expandafter \@secondoftwo
 \fi
}%
\providecommand \natexlab [1]{#1}%
\providecommand \enquote  [1]{``#1''}%
\providecommand \bibnamefont  [1]{#1}%
\providecommand \bibfnamefont [1]{#1}%
\providecommand \citenamefont [1]{#1}%
\providecommand \href@noop [0]{\@secondoftwo}%
\providecommand \href [0]{\begingroup \@sanitize@url \@href}%
\providecommand \@href[1]{\@@startlink{#1}\@@href}%
\providecommand \@@href[1]{\endgroup#1\@@endlink}%
\providecommand \@sanitize@url [0]{\catcode `\\12\catcode `\$12\catcode
  `\&12\catcode `\#12\catcode `\^12\catcode `\_12\catcode `\%12\relax}%
\providecommand \@@startlink[1]{}%
\providecommand \@@endlink[0]{}%
\providecommand \url  [0]{\begingroup\@sanitize@url \@url }%
\providecommand \@url [1]{\endgroup\@href {#1}{\urlprefix }}%
\providecommand \urlprefix  [0]{URL }%
\providecommand \Eprint [0]{\href }%
\providecommand \doibase [0]{http://dx.doi.org/}%
\providecommand \selectlanguage [0]{\@gobble}%
\providecommand \bibinfo  [0]{\@secondoftwo}%
\providecommand \bibfield  [0]{\@secondoftwo}%
\providecommand \translation [1]{[#1]}%
\providecommand \BibitemOpen [0]{}%
\providecommand \bibitemStop [0]{}%
\providecommand \bibitemNoStop [0]{.\EOS\space}%
\providecommand \EOS [0]{\spacefactor3000\relax}%
\providecommand \BibitemShut  [1]{\csname bibitem#1\endcsname}%
\let\auto@bib@innerbib\@empty
\bibitem [{\citenamefont {Nielsen}\ and\ \citenamefont
  {Chuang}(2011)}]{Nielsen2011}%
  \BibitemOpen
  \bibfield  {author} {\bibinfo {author} {\bibfnamefont {M.~A.}\ \bibnamefont
  {Nielsen}}\ and\ \bibinfo {author} {\bibfnamefont {I.~L.}\ \bibnamefont
  {Chuang}},\ }\href@noop {} {\emph {\bibinfo {title} {Quantum Computation and
  Quantum Information: 10th Anniversary Edition}}},\ \bibinfo {edition} {10th}\
  ed.\ (\bibinfo  {publisher} {Cambridge University Press},\ \bibinfo {address}
  {USA},\ \bibinfo {year} {2011})\BibitemShut {NoStop}%
\bibitem [{\citenamefont {Feynman}(1982)}]{feynman1982}%
  \BibitemOpen
  \bibfield  {author} {\bibinfo {author} {\bibfnamefont {R.~P.}\ \bibnamefont
  {Feynman}},\ }\bibfield  {title} {\enquote {\bibinfo {title} {Simulating
  physics with computers},}\ }\href@noop {} {\bibfield  {journal} {\bibinfo
  {journal} {International journal of theoretical physics}\ }\textbf {\bibinfo
  {volume} {21}},\ \bibinfo {pages} {467--488} (\bibinfo {year}
  {1982})}\BibitemShut {NoStop}%
\bibitem [{\citenamefont {Lidar}\ and\ \citenamefont
  {Brun}(2013)}]{lidar_brun_2013}%
  \BibitemOpen
  \bibfield  {author} {\bibinfo {author} {\bibfnamefont {D.~A.}\ \bibnamefont
  {Lidar}}\ and\ \bibinfo {author} {\bibfnamefont {T.~A.}\ \bibnamefont
  {Brun}},\ }\href {\doibase 10.1017/CBO9781139034807} {\emph {\bibinfo {title}
  {Quantum Error Correction}}}\ (\bibinfo  {publisher} {Cambridge University
  Press},\ \bibinfo {year} {2013})\BibitemShut {NoStop}%
\bibitem [{\citenamefont {Terhal}(2015)}]{TerhalRMP2015}%
  \BibitemOpen
  \bibfield  {author} {\bibinfo {author} {\bibfnamefont {B.~M.}\ \bibnamefont
  {Terhal}},\ }\bibfield  {title} {\enquote {\bibinfo {title} {Quantum error
  correction for quantum memories},}\ }\href {\doibase
  10.1103/RevModPhys.87.307} {\bibfield  {journal} {\bibinfo  {journal} {Rev.
  Mod. Phys.}\ }\textbf {\bibinfo {volume} {87}},\ \bibinfo {pages} {307--346}
  (\bibinfo {year} {2015})}\BibitemShut {NoStop}%
\bibitem [{\citenamefont {Wendin}(2017)}]{Wendin_2017}%
  \BibitemOpen
  \bibfield  {author} {\bibinfo {author} {\bibfnamefont {G.}~\bibnamefont
  {Wendin}},\ }\bibfield  {title} {\enquote {\bibinfo {title} {Quantum
  information processing with superconducting circuits: a review},}\ }\href
  {\doibase 10.1088/1361-6633/aa7e1a} {\bibfield  {journal} {\bibinfo
  {journal} {Reports on Progress in Physics}\ }\textbf {\bibinfo {volume}
  {80}},\ \bibinfo {pages} {106001} (\bibinfo {year} {2017})}\BibitemShut
  {NoStop}%
\bibitem [{\citenamefont {Bharti}\ \emph {et~al.}(2022)\citenamefont {Bharti},
  \citenamefont {Cervera-Lierta}, \citenamefont {Kyaw}, \citenamefont {Haug},
  \citenamefont {Alperin-Lea}, \citenamefont {Anand}, \citenamefont {Degroote},
  \citenamefont {Heimonen}, \citenamefont {Kottmann}, \citenamefont {Menke},
  \citenamefont {Mok}, \citenamefont {Sim}, \citenamefont {Kwek},\ and\
  \citenamefont {Aspuru-Guzik}}]{BhartiRMP2022}%
  \BibitemOpen
  \bibfield  {author} {\bibinfo {author} {\bibfnamefont {K.}~\bibnamefont
  {Bharti}}, \bibinfo {author} {\bibfnamefont {A.}~\bibnamefont
  {Cervera-Lierta}}, \bibinfo {author} {\bibfnamefont {T.~H.}\ \bibnamefont
  {Kyaw}}, \bibinfo {author} {\bibfnamefont {T.}~\bibnamefont {Haug}}, \bibinfo
  {author} {\bibfnamefont {S.}~\bibnamefont {Alperin-Lea}}, \bibinfo {author}
  {\bibfnamefont {A.}~\bibnamefont {Anand}}, \bibinfo {author} {\bibfnamefont
  {M.}~\bibnamefont {Degroote}}, \bibinfo {author} {\bibfnamefont
  {H.}~\bibnamefont {Heimonen}}, \bibinfo {author} {\bibfnamefont {J.~S.}\
  \bibnamefont {Kottmann}}, \bibinfo {author} {\bibfnamefont {T.}~\bibnamefont
  {Menke}}, \bibinfo {author} {\bibfnamefont {W.-K.}\ \bibnamefont {Mok}},
  \bibinfo {author} {\bibfnamefont {S.}~\bibnamefont {Sim}}, \bibinfo {author}
  {\bibfnamefont {L.-C.}\ \bibnamefont {Kwek}}, \ and\ \bibinfo {author}
  {\bibfnamefont {A.}~\bibnamefont {Aspuru-Guzik}},\ }\bibfield  {title}
  {\enquote {\bibinfo {title} {Noisy intermediate-scale quantum algorithms},}\
  }\href {\doibase 10.1103/RevModPhys.94.015004} {\bibfield  {journal}
  {\bibinfo  {journal} {Rev. Mod. Phys.}\ }\textbf {\bibinfo {volume} {94}},\
  \bibinfo {pages} {015004} (\bibinfo {year} {2022})}\BibitemShut {NoStop}%
\bibitem [{\citenamefont {Leymann}\ and\ \citenamefont
  {Barzen}(2020)}]{Leymann_2020}%
  \BibitemOpen
  \bibfield  {author} {\bibinfo {author} {\bibfnamefont {F.}~\bibnamefont
  {Leymann}}\ and\ \bibinfo {author} {\bibfnamefont {J.}~\bibnamefont
  {Barzen}},\ }\bibfield  {title} {\enquote {\bibinfo {title} {The bitter truth
  about gate-based quantum algorithms in the nisq era},}\ }\href {\doibase
  10.1088/2058-9565/abae7d} {\bibfield  {journal} {\bibinfo  {journal} {Quantum
  Science and Technology}\ }\textbf {\bibinfo {volume} {5}},\ \bibinfo {pages}
  {044007} (\bibinfo {year} {2020})}\BibitemShut {NoStop}%
\bibitem [{\citenamefont {Porter}\ and\ \citenamefont
  {Joseph}(2022)}]{Porter2022}%
  \BibitemOpen
  \bibfield  {author} {\bibinfo {author} {\bibfnamefont {M.~D.}\ \bibnamefont
  {Porter}}\ and\ \bibinfo {author} {\bibfnamefont {I.}~\bibnamefont
  {Joseph}},\ }\bibfield  {title} {\enquote {\bibinfo {title} {Observability of
  fidelity decay at the {L}yapunov rate in few-qubit quantum simulations},}\
  }\href {\doibase 10.22331/q-2022-09-08-799} {\bibfield  {journal} {\bibinfo
  {journal} {{Quantum}}\ }\textbf {\bibinfo {volume} {6}},\ \bibinfo {pages}
  {799} (\bibinfo {year} {2022})}\BibitemShut {NoStop}%
\bibitem [{\citenamefont {Kandala}\ \emph {et~al.}(2017)\citenamefont
  {Kandala}, \citenamefont {Mezzacapo}, \citenamefont {Temme}, \citenamefont
  {Takita}, \citenamefont {Brink}, \citenamefont {Chow},\ and\ \citenamefont
  {Gambetta}}]{Kandala2017}%
  \BibitemOpen
  \bibfield  {author} {\bibinfo {author} {\bibfnamefont {A.}~\bibnamefont
  {Kandala}}, \bibinfo {author} {\bibfnamefont {A.}~\bibnamefont {Mezzacapo}},
  \bibinfo {author} {\bibfnamefont {K.}~\bibnamefont {Temme}}, \bibinfo
  {author} {\bibfnamefont {M.}~\bibnamefont {Takita}}, \bibinfo {author}
  {\bibfnamefont {M.}~\bibnamefont {Brink}}, \bibinfo {author} {\bibfnamefont
  {J.~M.}\ \bibnamefont {Chow}}, \ and\ \bibinfo {author} {\bibfnamefont
  {J.~M.}\ \bibnamefont {Gambetta}},\ }\bibfield  {title} {\enquote {\bibinfo
  {title} {Hardware-efficient variational quantum eigensolver for small
  molecules and quantum magnets},}\ }\href {\doibase 10.1038/nature23879}
  {\bibfield  {journal} {\bibinfo  {journal} {Nature}\ }\textbf {\bibinfo
  {volume} {549}},\ \bibinfo {pages} {242--246} (\bibinfo {year}
  {2017})}\BibitemShut {NoStop}%
\bibitem [{\citenamefont {Aspuru-Guzik}\ \emph {et~al.}(2005)\citenamefont
  {Aspuru-Guzik}, \citenamefont {Dutoi}, \citenamefont {Love},\ and\
  \citenamefont {Head-Gordon}}]{Aspuru-Guzik-Mol-Science2005}%
  \BibitemOpen
  \bibfield  {author} {\bibinfo {author} {\bibfnamefont {A.}~\bibnamefont
  {Aspuru-Guzik}}, \bibinfo {author} {\bibfnamefont {A.~D.}\ \bibnamefont
  {Dutoi}}, \bibinfo {author} {\bibfnamefont {P.~J.}\ \bibnamefont {Love}}, \
  and\ \bibinfo {author} {\bibfnamefont {M.}~\bibnamefont {Head-Gordon}},\
  }\bibfield  {title} {\enquote {\bibinfo {title} {Simulated quantum
  computation of molecular energies},}\ }\href {\doibase
  10.1126/science.1113479} {\bibfield  {journal} {\bibinfo  {journal}
  {Science}\ }\textbf {\bibinfo {volume} {309}},\ \bibinfo {pages} {1704--1707}
  (\bibinfo {year} {2005})},\ \Eprint
  {http://arxiv.org/abs/https://www.science.org/doi/pdf/10.1126/science.1113479}
  {https://www.science.org/doi/pdf/10.1126/science.1113479} \BibitemShut
  {NoStop}%
\bibitem [{\citenamefont {Cerezo}\ \emph {et~al.}(2021)\citenamefont {Cerezo},
  \citenamefont {Arrasmith}, \citenamefont {Babbush}, \citenamefont {Benjamin},
  \citenamefont {Endo}, \citenamefont {Fujii}, \citenamefont {McClean},
  \citenamefont {Mitarai}, \citenamefont {Yuan}, \citenamefont {Cincio},\ and\
  \citenamefont {Coles}}]{Cerezo2021}%
  \BibitemOpen
  \bibfield  {author} {\bibinfo {author} {\bibfnamefont {M.}~\bibnamefont
  {Cerezo}}, \bibinfo {author} {\bibfnamefont {A.}~\bibnamefont {Arrasmith}},
  \bibinfo {author} {\bibfnamefont {R.}~\bibnamefont {Babbush}}, \bibinfo
  {author} {\bibfnamefont {S.~C.}\ \bibnamefont {Benjamin}}, \bibinfo {author}
  {\bibfnamefont {S.}~\bibnamefont {Endo}}, \bibinfo {author} {\bibfnamefont
  {K.}~\bibnamefont {Fujii}}, \bibinfo {author} {\bibfnamefont {J.~R.}\
  \bibnamefont {McClean}}, \bibinfo {author} {\bibfnamefont {K.}~\bibnamefont
  {Mitarai}}, \bibinfo {author} {\bibfnamefont {X.}~\bibnamefont {Yuan}},
  \bibinfo {author} {\bibfnamefont {L.}~\bibnamefont {Cincio}}, \ and\ \bibinfo
  {author} {\bibfnamefont {P.~J.}\ \bibnamefont {Coles}},\ }\bibfield  {title}
  {\enquote {\bibinfo {title} {Variational quantum algorithms},}\ }\href
  {\doibase 10.1038/s42254-021-00348-9} {\bibfield  {journal} {\bibinfo
  {journal} {Nature Reviews Physics}\ }\textbf {\bibinfo {volume} {3}},\
  \bibinfo {pages} {625--644} (\bibinfo {year} {2021})}\BibitemShut {NoStop}%
\bibitem [{\citenamefont {Arute}\ \emph {et~al.}(2019)\citenamefont {Arute},
  \citenamefont {Arya}, \citenamefont {Babbush}, \citenamefont {Bacon},
  \citenamefont {Bardin}, \citenamefont {Barends}, \citenamefont {Biswas},
  \citenamefont {Boixo}, \citenamefont {Brandao}, \citenamefont {Buell},
  \citenamefont {Burkett}, \citenamefont {Chen}, \citenamefont {Chen},
  \citenamefont {Chiaro}, \citenamefont {Collins}, \citenamefont {Courtney},
  \citenamefont {Dunsworth}, \citenamefont {Farhi}, \citenamefont {Foxen},
  \citenamefont {Fowler}, \citenamefont {Gidney}, \citenamefont {Giustina},
  \citenamefont {Graff}, \citenamefont {Guerin}, \citenamefont {Habegger},
  \citenamefont {Harrigan}, \citenamefont {Hartmann}, \citenamefont {Ho},
  \citenamefont {Hoffmann}, \citenamefont {Huang}, \citenamefont {Humble},
  \citenamefont {Isakov}, \citenamefont {Jeffrey}, \citenamefont {Jiang},
  \citenamefont {Kafri}, \citenamefont {Kechedzhi}, \citenamefont {Kelly},
  \citenamefont {Klimov}, \citenamefont {Knysh}, \citenamefont {Korotkov},
  \citenamefont {Kostritsa}, \citenamefont {Landhuis}, \citenamefont
  {Lindmark}, \citenamefont {Lucero}, \citenamefont {Lyakh}, \citenamefont
  {Mandr{\`a}}, \citenamefont {McClean}, \citenamefont {McEwen}, \citenamefont
  {Megrant}, \citenamefont {Mi}, \citenamefont {Michielsen}, \citenamefont
  {Mohseni}, \citenamefont {Mutus}, \citenamefont {Naaman}, \citenamefont
  {Neeley}, \citenamefont {Neill}, \citenamefont {Niu}, \citenamefont {Ostby},
  \citenamefont {Petukhov}, \citenamefont {Platt}, \citenamefont {Quintana},
  \citenamefont {Rieffel}, \citenamefont {Roushan}, \citenamefont {Rubin},
  \citenamefont {Sank}, \citenamefont {Satzinger}, \citenamefont {Smelyanskiy},
  \citenamefont {Sung}, \citenamefont {Trevithick}, \citenamefont
  {Vainsencher}, \citenamefont {Villalonga}, \citenamefont {White},
  \citenamefont {Yao}, \citenamefont {Yeh}, \citenamefont {Zalcman},
  \citenamefont {Neven},\ and\ \citenamefont {Martinis}}]{Arute2019}%
  \BibitemOpen
  \bibfield  {author} {\bibinfo {author} {\bibfnamefont {F.}~\bibnamefont
  {Arute}}, \bibinfo {author} {\bibfnamefont {K.}~\bibnamefont {Arya}},
  \bibinfo {author} {\bibfnamefont {R.}~\bibnamefont {Babbush}}, \bibinfo
  {author} {\bibfnamefont {D.}~\bibnamefont {Bacon}}, \bibinfo {author}
  {\bibfnamefont {J.~C.}\ \bibnamefont {Bardin}}, \bibinfo {author}
  {\bibfnamefont {R.}~\bibnamefont {Barends}}, \bibinfo {author} {\bibfnamefont
  {R.}~\bibnamefont {Biswas}}, \bibinfo {author} {\bibfnamefont
  {S.}~\bibnamefont {Boixo}}, \bibinfo {author} {\bibfnamefont {F.~G. S.~L.}\
  \bibnamefont {Brandao}}, \bibinfo {author} {\bibfnamefont {D.~A.}\
  \bibnamefont {Buell}}, \bibinfo {author} {\bibfnamefont {B.}~\bibnamefont
  {Burkett}}, \bibinfo {author} {\bibfnamefont {Y.}~\bibnamefont {Chen}},
  \bibinfo {author} {\bibfnamefont {Z.}~\bibnamefont {Chen}}, \bibinfo {author}
  {\bibfnamefont {B.}~\bibnamefont {Chiaro}}, \bibinfo {author} {\bibfnamefont
  {R.}~\bibnamefont {Collins}}, \bibinfo {author} {\bibfnamefont
  {W.}~\bibnamefont {Courtney}}, \bibinfo {author} {\bibfnamefont
  {A.}~\bibnamefont {Dunsworth}}, \bibinfo {author} {\bibfnamefont
  {E.}~\bibnamefont {Farhi}}, \bibinfo {author} {\bibfnamefont
  {B.}~\bibnamefont {Foxen}}, \bibinfo {author} {\bibfnamefont
  {A.}~\bibnamefont {Fowler}}, \bibinfo {author} {\bibfnamefont
  {C.}~\bibnamefont {Gidney}}, \bibinfo {author} {\bibfnamefont
  {M.}~\bibnamefont {Giustina}}, \bibinfo {author} {\bibfnamefont
  {R.}~\bibnamefont {Graff}}, \bibinfo {author} {\bibfnamefont
  {K.}~\bibnamefont {Guerin}}, \bibinfo {author} {\bibfnamefont
  {S.}~\bibnamefont {Habegger}}, \bibinfo {author} {\bibfnamefont {M.~P.}\
  \bibnamefont {Harrigan}}, \bibinfo {author} {\bibfnamefont {M.~J.}\
  \bibnamefont {Hartmann}}, \bibinfo {author} {\bibfnamefont {A.}~\bibnamefont
  {Ho}}, \bibinfo {author} {\bibfnamefont {M.}~\bibnamefont {Hoffmann}},
  \bibinfo {author} {\bibfnamefont {T.}~\bibnamefont {Huang}}, \bibinfo
  {author} {\bibfnamefont {T.~S.}\ \bibnamefont {Humble}}, \bibinfo {author}
  {\bibfnamefont {S.~V.}\ \bibnamefont {Isakov}}, \bibinfo {author}
  {\bibfnamefont {E.}~\bibnamefont {Jeffrey}}, \bibinfo {author} {\bibfnamefont
  {Z.}~\bibnamefont {Jiang}}, \bibinfo {author} {\bibfnamefont
  {D.}~\bibnamefont {Kafri}}, \bibinfo {author} {\bibfnamefont
  {K.}~\bibnamefont {Kechedzhi}}, \bibinfo {author} {\bibfnamefont
  {J.}~\bibnamefont {Kelly}}, \bibinfo {author} {\bibfnamefont {P.~V.}\
  \bibnamefont {Klimov}}, \bibinfo {author} {\bibfnamefont {S.}~\bibnamefont
  {Knysh}}, \bibinfo {author} {\bibfnamefont {A.}~\bibnamefont {Korotkov}},
  \bibinfo {author} {\bibfnamefont {F.}~\bibnamefont {Kostritsa}}, \bibinfo
  {author} {\bibfnamefont {D.}~\bibnamefont {Landhuis}}, \bibinfo {author}
  {\bibfnamefont {M.}~\bibnamefont {Lindmark}}, \bibinfo {author}
  {\bibfnamefont {E.}~\bibnamefont {Lucero}}, \bibinfo {author} {\bibfnamefont
  {D.}~\bibnamefont {Lyakh}}, \bibinfo {author} {\bibfnamefont
  {S.}~\bibnamefont {Mandr{\`a}}}, \bibinfo {author} {\bibfnamefont {J.~R.}\
  \bibnamefont {McClean}}, \bibinfo {author} {\bibfnamefont {M.}~\bibnamefont
  {McEwen}}, \bibinfo {author} {\bibfnamefont {A.}~\bibnamefont {Megrant}},
  \bibinfo {author} {\bibfnamefont {X.}~\bibnamefont {Mi}}, \bibinfo {author}
  {\bibfnamefont {K.}~\bibnamefont {Michielsen}}, \bibinfo {author}
  {\bibfnamefont {M.}~\bibnamefont {Mohseni}}, \bibinfo {author} {\bibfnamefont
  {J.}~\bibnamefont {Mutus}}, \bibinfo {author} {\bibfnamefont
  {O.}~\bibnamefont {Naaman}}, \bibinfo {author} {\bibfnamefont
  {M.}~\bibnamefont {Neeley}}, \bibinfo {author} {\bibfnamefont
  {C.}~\bibnamefont {Neill}}, \bibinfo {author} {\bibfnamefont {M.~Y.}\
  \bibnamefont {Niu}}, \bibinfo {author} {\bibfnamefont {E.}~\bibnamefont
  {Ostby}}, \bibinfo {author} {\bibfnamefont {A.}~\bibnamefont {Petukhov}},
  \bibinfo {author} {\bibfnamefont {J.~C.}\ \bibnamefont {Platt}}, \bibinfo
  {author} {\bibfnamefont {C.}~\bibnamefont {Quintana}}, \bibinfo {author}
  {\bibfnamefont {E.~G.}\ \bibnamefont {Rieffel}}, \bibinfo {author}
  {\bibfnamefont {P.}~\bibnamefont {Roushan}}, \bibinfo {author} {\bibfnamefont
  {N.~C.}\ \bibnamefont {Rubin}}, \bibinfo {author} {\bibfnamefont
  {D.}~\bibnamefont {Sank}}, \bibinfo {author} {\bibfnamefont {K.~J.}\
  \bibnamefont {Satzinger}}, \bibinfo {author} {\bibfnamefont {V.}~\bibnamefont
  {Smelyanskiy}}, \bibinfo {author} {\bibfnamefont {K.~J.}\ \bibnamefont
  {Sung}}, \bibinfo {author} {\bibfnamefont {M.~D.}\ \bibnamefont
  {Trevithick}}, \bibinfo {author} {\bibfnamefont {A.}~\bibnamefont
  {Vainsencher}}, \bibinfo {author} {\bibfnamefont {B.}~\bibnamefont
  {Villalonga}}, \bibinfo {author} {\bibfnamefont {T.}~\bibnamefont {White}},
  \bibinfo {author} {\bibfnamefont {Z.~J.}\ \bibnamefont {Yao}}, \bibinfo
  {author} {\bibfnamefont {P.}~\bibnamefont {Yeh}}, \bibinfo {author}
  {\bibfnamefont {A.}~\bibnamefont {Zalcman}}, \bibinfo {author} {\bibfnamefont
  {H.}~\bibnamefont {Neven}}, \ and\ \bibinfo {author} {\bibfnamefont {J.~M.}\
  \bibnamefont {Martinis}},\ }\bibfield  {title} {\enquote {\bibinfo {title}
  {Quantum supremacy using a programmable superconducting processor},}\ }\href
  {\doibase 10.1038/s41586-019-1666-5} {\bibfield  {journal} {\bibinfo
  {journal} {Nature}\ }\textbf {\bibinfo {volume} {574}},\ \bibinfo {pages}
  {505--510} (\bibinfo {year} {2019})}\BibitemShut {NoStop}%
\bibitem [{\citenamefont {Preskill}(2018)}]{Preskill2018}%
  \BibitemOpen
  \bibfield  {author} {\bibinfo {author} {\bibfnamefont {J.}~\bibnamefont
  {Preskill}},\ }\bibfield  {title} {\enquote {\bibinfo {title} {Quantum
  {C}omputing in the {NISQ} era and beyond},}\ }\href {\doibase
  10.22331/q-2018-08-06-79} {\bibfield  {journal} {\bibinfo  {journal}
  {{Quantum}}\ }\textbf {\bibinfo {volume} {2}},\ \bibinfo {pages} {79}
  (\bibinfo {year} {2018})}\BibitemShut {NoStop}%
\bibitem [{\citenamefont {Xiao}, \citenamefont {Freericks},\ and\ \citenamefont
  {Kemper}(2021)}]{Xiao2021}%
  \BibitemOpen
  \bibfield  {author} {\bibinfo {author} {\bibfnamefont {X.}~\bibnamefont
  {Xiao}}, \bibinfo {author} {\bibfnamefont {J.~K.}\ \bibnamefont {Freericks}},
  \ and\ \bibinfo {author} {\bibfnamefont {A.~F.}\ \bibnamefont {Kemper}},\
  }\bibfield  {title} {\enquote {\bibinfo {title} {Determining quantum phase
  diagrams of topological {K}itaev-inspired models on {NISQ} quantum
  hardware},}\ }\href {\doibase 10.22331/q-2021-09-28-553} {\bibfield
  {journal} {\bibinfo  {journal} {{Quantum}}\ }\textbf {\bibinfo {volume}
  {5}},\ \bibinfo {pages} {553} (\bibinfo {year} {2021})}\BibitemShut {NoStop}%
\bibitem [{\citenamefont {Dalzell}\ \emph {et~al.}(2020)\citenamefont
  {Dalzell}, \citenamefont {Harrow}, \citenamefont {Koh},\ and\ \citenamefont
  {La~Placa}}]{Dalzell2020}%
  \BibitemOpen
  \bibfield  {author} {\bibinfo {author} {\bibfnamefont {A.~M.}\ \bibnamefont
  {Dalzell}}, \bibinfo {author} {\bibfnamefont {A.~W.}\ \bibnamefont {Harrow}},
  \bibinfo {author} {\bibfnamefont {D.~E.}\ \bibnamefont {Koh}}, \ and\
  \bibinfo {author} {\bibfnamefont {R.~L.}\ \bibnamefont {La~Placa}},\
  }\bibfield  {title} {\enquote {\bibinfo {title} {How many qubits are needed
  for quantum computational supremacy?}}\ }\href {\doibase
  10.22331/q-2020-05-11-264} {\bibfield  {journal} {\bibinfo  {journal}
  {{Quantum}}\ }\textbf {\bibinfo {volume} {4}},\ \bibinfo {pages} {264}
  (\bibinfo {year} {2020})}\BibitemShut {NoStop}%
\bibitem [{\citenamefont {Georgopoulos}, \citenamefont {Emary},\ and\
  \citenamefont {Zuliani}(2021)}]{georgopoulos2021modeling}%
  \BibitemOpen
  \bibfield  {author} {\bibinfo {author} {\bibfnamefont {K.}~\bibnamefont
  {Georgopoulos}}, \bibinfo {author} {\bibfnamefont {C.}~\bibnamefont {Emary}},
  \ and\ \bibinfo {author} {\bibfnamefont {P.}~\bibnamefont {Zuliani}},\
  }\bibfield  {title} {\enquote {\bibinfo {title} {Modeling and simulating the
  noisy behavior of near-term quantum computers},}\ }\href@noop {} {\bibfield
  {journal} {\bibinfo  {journal} {Physical Review A}\ }\textbf {\bibinfo
  {volume} {104}},\ \bibinfo {pages} {062432} (\bibinfo {year}
  {2021})}\BibitemShut {NoStop}%
\bibitem [{\citenamefont {Patel}\ \emph {et~al.}(2020)\citenamefont {Patel},
  \citenamefont {Potharaju}, \citenamefont {Li}, \citenamefont {Roy},\ and\
  \citenamefont {Tiwari}}]{patel2020experimental}%
  \BibitemOpen
  \bibfield  {author} {\bibinfo {author} {\bibfnamefont {T.}~\bibnamefont
  {Patel}}, \bibinfo {author} {\bibfnamefont {A.}~\bibnamefont {Potharaju}},
  \bibinfo {author} {\bibfnamefont {B.}~\bibnamefont {Li}}, \bibinfo {author}
  {\bibfnamefont {R.~B.}\ \bibnamefont {Roy}}, \ and\ \bibinfo {author}
  {\bibfnamefont {D.}~\bibnamefont {Tiwari}},\ }\bibfield  {title} {\enquote
  {\bibinfo {title} {Experimental evaluation of nisq quantum computers: error
  measurement, characterization, and implications},}\ }in\ \href@noop {} {\emph
  {\bibinfo {booktitle} {SC20: International Conference for High Performance
  Computing, Networking, Storage and Analysis}}}\ (\bibinfo {organization}
  {IEEE},\ \bibinfo {year} {2020})\ pp.\ \bibinfo {pages} {1--15}\BibitemShut
  {NoStop}%
\bibitem [{\citenamefont {Nation}\ \emph {et~al.}(2021)\citenamefont {Nation},
  \citenamefont {Kang}, \citenamefont {Sundaresan},\ and\ \citenamefont
  {Gambetta}}]{nation2021scalable}%
  \BibitemOpen
  \bibfield  {author} {\bibinfo {author} {\bibfnamefont {P.~D.}\ \bibnamefont
  {Nation}}, \bibinfo {author} {\bibfnamefont {H.}~\bibnamefont {Kang}},
  \bibinfo {author} {\bibfnamefont {N.}~\bibnamefont {Sundaresan}}, \ and\
  \bibinfo {author} {\bibfnamefont {J.~M.}\ \bibnamefont {Gambetta}},\
  }\bibfield  {title} {\enquote {\bibinfo {title} {Scalable mitigation of
  measurement errors on quantum computers},}\ }\href@noop {} {\bibfield
  {journal} {\bibinfo  {journal} {PRX Quantum}\ }\textbf {\bibinfo {volume}
  {2}},\ \bibinfo {pages} {040326} (\bibinfo {year} {2021})}\BibitemShut
  {NoStop}%
\bibitem [{\citenamefont {Weidenfeller}\ \emph {et~al.}(2022)\citenamefont
  {Weidenfeller}, \citenamefont {Valor}, \citenamefont {Gacon}, \citenamefont
  {Tornow}, \citenamefont {Bello}, \citenamefont {Woerner},\ and\ \citenamefont
  {Egger}}]{Weidenfeller2022}%
  \BibitemOpen
  \bibfield  {author} {\bibinfo {author} {\bibfnamefont {J.}~\bibnamefont
  {Weidenfeller}}, \bibinfo {author} {\bibfnamefont {L.~C.}\ \bibnamefont
  {Valor}}, \bibinfo {author} {\bibfnamefont {J.}~\bibnamefont {Gacon}},
  \bibinfo {author} {\bibfnamefont {C.}~\bibnamefont {Tornow}}, \bibinfo
  {author} {\bibfnamefont {L.}~\bibnamefont {Bello}}, \bibinfo {author}
  {\bibfnamefont {S.}~\bibnamefont {Woerner}}, \ and\ \bibinfo {author}
  {\bibfnamefont {D.~J.}\ \bibnamefont {Egger}},\ }\bibfield  {title} {\enquote
  {\bibinfo {title} {Scaling of the quantum approximate optimization algorithm
  on superconducting qubit based hardware},}\ }\href {\doibase
  10.22331/q-2022-12-07-870} {\bibfield  {journal} {\bibinfo  {journal}
  {{Quantum}}\ }\textbf {\bibinfo {volume} {6}},\ \bibinfo {pages} {870}
  (\bibinfo {year} {2022})}\BibitemShut {NoStop}%
\bibitem [{\citenamefont {Setiawan}\ \emph {et~al.}(2021)\citenamefont
  {Setiawan}, \citenamefont {Groszkowski}, \citenamefont {Ribeiro},\ and\
  \citenamefont {Clerk}}]{SetiawanPRX2021}%
  \BibitemOpen
  \bibfield  {author} {\bibinfo {author} {\bibfnamefont {F.}~\bibnamefont
  {Setiawan}}, \bibinfo {author} {\bibfnamefont {P.}~\bibnamefont
  {Groszkowski}}, \bibinfo {author} {\bibfnamefont {H.}~\bibnamefont
  {Ribeiro}}, \ and\ \bibinfo {author} {\bibfnamefont {A.~A.}\ \bibnamefont
  {Clerk}},\ }\bibfield  {title} {\enquote {\bibinfo {title} {Analytic design
  of accelerated adiabatic gates in realistic qubits: General theory and
  applications to superconducting circuits},}\ }\href {\doibase
  10.1103/PRXQuantum.2.030306} {\bibfield  {journal} {\bibinfo  {journal} {PRX
  Quantum}\ }\textbf {\bibinfo {volume} {2}},\ \bibinfo {pages} {030306}
  (\bibinfo {year} {2021})}\BibitemShut {NoStop}%
\bibitem [{\citenamefont {Wu}\ \emph {et~al.}(2021)\citenamefont {Wu},
  \citenamefont {Bao}, \citenamefont {Cao}, \citenamefont {Chen}, \citenamefont
  {Chen}, \citenamefont {Chen}, \citenamefont {Chung}, \citenamefont {Deng},
  \citenamefont {Du}, \citenamefont {Fan}, \citenamefont {Gong}, \citenamefont
  {Guo}, \citenamefont {Guo}, \citenamefont {Guo}, \citenamefont {Han},
  \citenamefont {Hong}, \citenamefont {Huang}, \citenamefont {Huo},
  \citenamefont {Li}, \citenamefont {Li}, \citenamefont {Li}, \citenamefont
  {Li}, \citenamefont {Liang}, \citenamefont {Lin}, \citenamefont {Lin},
  \citenamefont {Qian}, \citenamefont {Qiao}, \citenamefont {Rong},
  \citenamefont {Su}, \citenamefont {Sun}, \citenamefont {Wang}, \citenamefont
  {Wang}, \citenamefont {Wu}, \citenamefont {Xu}, \citenamefont {Yan},
  \citenamefont {Yang}, \citenamefont {Yang}, \citenamefont {Ye}, \citenamefont
  {Yin}, \citenamefont {Ying}, \citenamefont {Yu}, \citenamefont {Zha},
  \citenamefont {Zhang}, \citenamefont {Zhang}, \citenamefont {Zhang},
  \citenamefont {Zhang}, \citenamefont {Zhao}, \citenamefont {Zhao},
  \citenamefont {Zhou}, \citenamefont {Zhu}, \citenamefont {Lu}, \citenamefont
  {Peng}, \citenamefont {Zhu},\ and\ \citenamefont {Pan}}]{WuPRL2021}%
  \BibitemOpen
  \bibfield  {author} {\bibinfo {author} {\bibfnamefont {Y.}~\bibnamefont
  {Wu}}, \bibinfo {author} {\bibfnamefont {W.-S.}\ \bibnamefont {Bao}},
  \bibinfo {author} {\bibfnamefont {S.}~\bibnamefont {Cao}}, \bibinfo {author}
  {\bibfnamefont {F.}~\bibnamefont {Chen}}, \bibinfo {author} {\bibfnamefont
  {M.-C.}\ \bibnamefont {Chen}}, \bibinfo {author} {\bibfnamefont
  {X.}~\bibnamefont {Chen}}, \bibinfo {author} {\bibfnamefont {T.-H.}\
  \bibnamefont {Chung}}, \bibinfo {author} {\bibfnamefont {H.}~\bibnamefont
  {Deng}}, \bibinfo {author} {\bibfnamefont {Y.}~\bibnamefont {Du}}, \bibinfo
  {author} {\bibfnamefont {D.}~\bibnamefont {Fan}}, \bibinfo {author}
  {\bibfnamefont {M.}~\bibnamefont {Gong}}, \bibinfo {author} {\bibfnamefont
  {C.}~\bibnamefont {Guo}}, \bibinfo {author} {\bibfnamefont {C.}~\bibnamefont
  {Guo}}, \bibinfo {author} {\bibfnamefont {S.}~\bibnamefont {Guo}}, \bibinfo
  {author} {\bibfnamefont {L.}~\bibnamefont {Han}}, \bibinfo {author}
  {\bibfnamefont {L.}~\bibnamefont {Hong}}, \bibinfo {author} {\bibfnamefont
  {H.-L.}\ \bibnamefont {Huang}}, \bibinfo {author} {\bibfnamefont {Y.-H.}\
  \bibnamefont {Huo}}, \bibinfo {author} {\bibfnamefont {L.}~\bibnamefont
  {Li}}, \bibinfo {author} {\bibfnamefont {N.}~\bibnamefont {Li}}, \bibinfo
  {author} {\bibfnamefont {S.}~\bibnamefont {Li}}, \bibinfo {author}
  {\bibfnamefont {Y.}~\bibnamefont {Li}}, \bibinfo {author} {\bibfnamefont
  {F.}~\bibnamefont {Liang}}, \bibinfo {author} {\bibfnamefont
  {C.}~\bibnamefont {Lin}}, \bibinfo {author} {\bibfnamefont {J.}~\bibnamefont
  {Lin}}, \bibinfo {author} {\bibfnamefont {H.}~\bibnamefont {Qian}}, \bibinfo
  {author} {\bibfnamefont {D.}~\bibnamefont {Qiao}}, \bibinfo {author}
  {\bibfnamefont {H.}~\bibnamefont {Rong}}, \bibinfo {author} {\bibfnamefont
  {H.}~\bibnamefont {Su}}, \bibinfo {author} {\bibfnamefont {L.}~\bibnamefont
  {Sun}}, \bibinfo {author} {\bibfnamefont {L.}~\bibnamefont {Wang}}, \bibinfo
  {author} {\bibfnamefont {S.}~\bibnamefont {Wang}}, \bibinfo {author}
  {\bibfnamefont {D.}~\bibnamefont {Wu}}, \bibinfo {author} {\bibfnamefont
  {Y.}~\bibnamefont {Xu}}, \bibinfo {author} {\bibfnamefont {K.}~\bibnamefont
  {Yan}}, \bibinfo {author} {\bibfnamefont {W.}~\bibnamefont {Yang}}, \bibinfo
  {author} {\bibfnamefont {Y.}~\bibnamefont {Yang}}, \bibinfo {author}
  {\bibfnamefont {Y.}~\bibnamefont {Ye}}, \bibinfo {author} {\bibfnamefont
  {J.}~\bibnamefont {Yin}}, \bibinfo {author} {\bibfnamefont {C.}~\bibnamefont
  {Ying}}, \bibinfo {author} {\bibfnamefont {J.}~\bibnamefont {Yu}}, \bibinfo
  {author} {\bibfnamefont {C.}~\bibnamefont {Zha}}, \bibinfo {author}
  {\bibfnamefont {C.}~\bibnamefont {Zhang}}, \bibinfo {author} {\bibfnamefont
  {H.}~\bibnamefont {Zhang}}, \bibinfo {author} {\bibfnamefont
  {K.}~\bibnamefont {Zhang}}, \bibinfo {author} {\bibfnamefont
  {Y.}~\bibnamefont {Zhang}}, \bibinfo {author} {\bibfnamefont
  {H.}~\bibnamefont {Zhao}}, \bibinfo {author} {\bibfnamefont {Y.}~\bibnamefont
  {Zhao}}, \bibinfo {author} {\bibfnamefont {L.}~\bibnamefont {Zhou}}, \bibinfo
  {author} {\bibfnamefont {Q.}~\bibnamefont {Zhu}}, \bibinfo {author}
  {\bibfnamefont {C.-Y.}\ \bibnamefont {Lu}}, \bibinfo {author} {\bibfnamefont
  {C.-Z.}\ \bibnamefont {Peng}}, \bibinfo {author} {\bibfnamefont
  {X.}~\bibnamefont {Zhu}}, \ and\ \bibinfo {author} {\bibfnamefont {J.-W.}\
  \bibnamefont {Pan}},\ }\bibfield  {title} {\enquote {\bibinfo {title} {Strong
  quantum computational advantage using a superconducting quantum processor},}\
  }\href {\doibase 10.1103/PhysRevLett.127.180501} {\bibfield  {journal}
  {\bibinfo  {journal} {Phys. Rev. Lett.}\ }\textbf {\bibinfo {volume} {127}},\
  \bibinfo {pages} {180501} (\bibinfo {year} {2021})}\BibitemShut {NoStop}%
\bibitem [{\citenamefont {Headley}\ \emph {et~al.}(2022)\citenamefont
  {Headley}, \citenamefont {M\"uller}, \citenamefont {Martin}, \citenamefont
  {Solano}, \citenamefont {Sanz},\ and\ \citenamefont
  {Wilhelm}}]{HeadleyPRA2022}%
  \BibitemOpen
  \bibfield  {author} {\bibinfo {author} {\bibfnamefont {D.}~\bibnamefont
  {Headley}}, \bibinfo {author} {\bibfnamefont {T.}~\bibnamefont {M\"uller}},
  \bibinfo {author} {\bibfnamefont {A.}~\bibnamefont {Martin}}, \bibinfo
  {author} {\bibfnamefont {E.}~\bibnamefont {Solano}}, \bibinfo {author}
  {\bibfnamefont {M.}~\bibnamefont {Sanz}}, \ and\ \bibinfo {author}
  {\bibfnamefont {F.~K.}\ \bibnamefont {Wilhelm}},\ }\bibfield  {title}
  {\enquote {\bibinfo {title} {Approximating the quantum approximate
  optimization algorithm with digital-analog interactions},}\ }\href {\doibase
  10.1103/PhysRevA.106.042446} {\bibfield  {journal} {\bibinfo  {journal}
  {Phys. Rev. A}\ }\textbf {\bibinfo {volume} {106}},\ \bibinfo {pages}
  {042446} (\bibinfo {year} {2022})}\BibitemShut {NoStop}%
\bibitem [{\citenamefont {Koch}\ \emph {et~al.}(2007)\citenamefont {Koch},
  \citenamefont {Yu}, \citenamefont {Gambetta}, \citenamefont {Houck},
  \citenamefont {Schuster}, \citenamefont {Majer}, \citenamefont {Blais},
  \citenamefont {Devoret}, \citenamefont {Girvin},\ and\ \citenamefont
  {Schoelkopf}}]{transmon}%
  \BibitemOpen
  \bibfield  {author} {\bibinfo {author} {\bibfnamefont {J.}~\bibnamefont
  {Koch}}, \bibinfo {author} {\bibfnamefont {T.~M.}\ \bibnamefont {Yu}},
  \bibinfo {author} {\bibfnamefont {J.}~\bibnamefont {Gambetta}}, \bibinfo
  {author} {\bibfnamefont {A.~A.}\ \bibnamefont {Houck}}, \bibinfo {author}
  {\bibfnamefont {D.~I.}\ \bibnamefont {Schuster}}, \bibinfo {author}
  {\bibfnamefont {J.}~\bibnamefont {Majer}}, \bibinfo {author} {\bibfnamefont
  {A.}~\bibnamefont {Blais}}, \bibinfo {author} {\bibfnamefont {M.~H.}\
  \bibnamefont {Devoret}}, \bibinfo {author} {\bibfnamefont {S.~M.}\
  \bibnamefont {Girvin}}, \ and\ \bibinfo {author} {\bibfnamefont {R.~J.}\
  \bibnamefont {Schoelkopf}},\ }\bibfield  {title} {\enquote {\bibinfo {title}
  {Charge-insensitive qubit design derived from the cooper pair box},}\ }\href
  {\doibase 10.1103/PhysRevA.76.042319} {\bibfield  {journal} {\bibinfo
  {journal} {Phys. Rev. A}\ }\textbf {\bibinfo {volume} {76}},\ \bibinfo
  {pages} {042319} (\bibinfo {year} {2007})}\BibitemShut {NoStop}%
\bibitem [{\citenamefont {Kim}\ \emph {et~al.}(2023)\citenamefont {Kim},
  \citenamefont {Eddins}, \citenamefont {Anand}, \citenamefont {Wei},
  \citenamefont {van~den Berg}, \citenamefont {Rosenblatt}, \citenamefont
  {Nayfeh}, \citenamefont {Wu}, \citenamefont {Zaletel}, \citenamefont
  {Temme},\ and\ \citenamefont {Kandala}}]{natureIBM2023}%
  \BibitemOpen
  \bibfield  {author} {\bibinfo {author} {\bibfnamefont {Y.}~\bibnamefont
  {Kim}}, \bibinfo {author} {\bibfnamefont {A.}~\bibnamefont {Eddins}},
  \bibinfo {author} {\bibfnamefont {S.}~\bibnamefont {Anand}}, \bibinfo
  {author} {\bibfnamefont {K.~X.}\ \bibnamefont {Wei}}, \bibinfo {author}
  {\bibfnamefont {E.}~\bibnamefont {van~den Berg}}, \bibinfo {author}
  {\bibfnamefont {S.}~\bibnamefont {Rosenblatt}}, \bibinfo {author}
  {\bibfnamefont {H.}~\bibnamefont {Nayfeh}}, \bibinfo {author} {\bibfnamefont
  {Y.}~\bibnamefont {Wu}}, \bibinfo {author} {\bibfnamefont {M.}~\bibnamefont
  {Zaletel}}, \bibinfo {author} {\bibfnamefont {K.}~\bibnamefont {Temme}}, \
  and\ \bibinfo {author} {\bibfnamefont {A.}~\bibnamefont {Kandala}},\
  }\bibfield  {title} {\enquote {\bibinfo {title} {Evidence for the utility of
  quantum computing before fault tolerance},}\ }\href {\doibase
  10.1038/s41586-023-06096-3} {\bibfield  {journal} {\bibinfo  {journal}
  {Nature}\ }\textbf {\bibinfo {volume} {618}},\ \bibinfo {pages} {500}
  (\bibinfo {year} {2023})}\BibitemShut {NoStop}%
\bibitem [{\citenamefont {AI}\ and\ \citenamefont
  {Collaborators}(2024)}]{natureGoogle2024}%
  \BibitemOpen
  \bibfield  {author} {\bibinfo {author} {\bibfnamefont {G.~Q.}\ \bibnamefont
  {AI}}\ and\ \bibinfo {author} {\bibnamefont {Collaborators}},\ }\bibfield
  {title} {\enquote {\bibinfo {title} {Quantum error correction below the
  surface code threshold},}\ }\href {\doibase 10.1038/s41586-024-08449-y}
  {\bibfield  {journal} {\bibinfo  {journal} {Nature}\ } (\bibinfo {year}
  {2024}),\ 10.1038/s41586-024-08449-y}\BibitemShut {NoStop}%
\bibitem [{\citenamefont {Aseguinolaza}\ \emph {et~al.}(2024)\citenamefont
  {Aseguinolaza}, \citenamefont {Sobrino}, \citenamefont {Sobrino},
  \citenamefont {Jornet-Somoza},\ and\ \citenamefont
  {Borge}}]{Aseginolaza2024}%
  \BibitemOpen
  \bibfield  {author} {\bibinfo {author} {\bibfnamefont {U.}~\bibnamefont
  {Aseguinolaza}}, \bibinfo {author} {\bibfnamefont {N.}~\bibnamefont
  {Sobrino}}, \bibinfo {author} {\bibfnamefont {G.}~\bibnamefont {Sobrino}},
  \bibinfo {author} {\bibfnamefont {J.}~\bibnamefont {Jornet-Somoza}}, \ and\
  \bibinfo {author} {\bibfnamefont {J.}~\bibnamefont {Borge}},\ }\bibfield
  {title} {\enquote {\bibinfo {title} {Error estimation in current noisy
  quantum computers},}\ }\href {\doibase 10.1007/s11128-024-04384-z} {\bibfield
   {journal} {\bibinfo  {journal} {Quantum Information Processing}\ }\textbf
  {\bibinfo {volume} {23}},\ \bibinfo {pages} {181 5} (\bibinfo {year}
  {2024})}\BibitemShut {NoStop}%
\bibitem [{\citenamefont {Brody}\ and\ \citenamefont
  {Sykes}(2025)}]{brody2025grover}%
  \BibitemOpen
  \bibfield  {author} {\bibinfo {author} {\bibfnamefont {J.}~\bibnamefont
  {Brody}}\ and\ \bibinfo {author} {\bibfnamefont {G.-C.~W.}\ \bibnamefont
  {Sykes}},\ }\bibfield  {title} {\enquote {\bibinfo {title} {Grover’s search
  algorithm: An approachable application of quantum computing},}\ }\href@noop
  {} {\bibfield  {journal} {\bibinfo  {journal} {The Physics Teacher}\ }\textbf
  {\bibinfo {volume} {63}},\ \bibinfo {pages} {23--25} (\bibinfo {year}
  {2025})}\BibitemShut {NoStop}%
\bibitem [{\citenamefont {Gardeazabal-Gutierrez}, \citenamefont
  {Terres-Escudero},\ and\ \citenamefont
  {Bringas}(2024)}]{gardeazabal2024machine}%
  \BibitemOpen
  \bibfield  {author} {\bibinfo {author} {\bibfnamefont {J.}~\bibnamefont
  {Gardeazabal-Gutierrez}}, \bibinfo {author} {\bibfnamefont {E.~B.}\
  \bibnamefont {Terres-Escudero}}, \ and\ \bibinfo {author} {\bibfnamefont
  {P.~G.}\ \bibnamefont {Bringas}},\ }\bibfield  {title} {\enquote {\bibinfo
  {title} {Machine learning methods as robust quantum noise estimators},}\ }in\
  \href@noop {} {\emph {\bibinfo {booktitle} {International Conference on
  Hybrid Artificial Intelligence Systems}}}\ (\bibinfo {organization}
  {Springer},\ \bibinfo {year} {2024})\ pp.\ \bibinfo {pages}
  {238--247}\BibitemShut {NoStop}%
\bibitem [{\citenamefont {Antero}\ \emph {et~al.}(2025)\citenamefont {Antero},
  \citenamefont {Sierra}, \citenamefont {O{\~n}ativia}, \citenamefont {Ruiz},\
  and\ \citenamefont {Osaba}}]{antero2025robot}%
  \BibitemOpen
  \bibfield  {author} {\bibinfo {author} {\bibfnamefont {U.}~\bibnamefont
  {Antero}}, \bibinfo {author} {\bibfnamefont {B.}~\bibnamefont {Sierra}},
  \bibinfo {author} {\bibfnamefont {J.}~\bibnamefont {O{\~n}ativia}}, \bibinfo
  {author} {\bibfnamefont {A.}~\bibnamefont {Ruiz}}, \ and\ \bibinfo {author}
  {\bibfnamefont {E.}~\bibnamefont {Osaba}},\ }\bibfield  {title} {\enquote
  {\bibinfo {title} {Robot localization aided by quantum algorithms},}\
  }\href@noop {} {\bibfield  {journal} {\bibinfo  {journal} {arXiv preprint
  arXiv:2502.00077}\ } (\bibinfo {year} {2025})}\BibitemShut {NoStop}%
\bibitem [{\citenamefont {de~Keijzer}\ \emph {et~al.}(2025)\citenamefont
  {de~Keijzer}, \citenamefont {Visser}, \citenamefont {Tse},\ and\
  \citenamefont {Kokkelmans}}]{de2025fidelity}%
  \BibitemOpen
  \bibfield  {author} {\bibinfo {author} {\bibfnamefont {R.}~\bibnamefont
  {de~Keijzer}}, \bibinfo {author} {\bibfnamefont {L.}~\bibnamefont {Visser}},
  \bibinfo {author} {\bibfnamefont {O.}~\bibnamefont {Tse}}, \ and\ \bibinfo
  {author} {\bibfnamefont {S.}~\bibnamefont {Kokkelmans}},\ }\bibfield  {title}
  {\enquote {\bibinfo {title} {Fidelity-enhanced variational quantum optimal
  control},}\ }\href@noop {} {\bibfield  {journal} {\bibinfo  {journal} {arXiv
  preprint arXiv:2501.17692}\ } (\bibinfo {year} {2025})}\BibitemShut {NoStop}%
\bibitem [{\citenamefont {Cai}\ \emph {et~al.}(2023)\citenamefont {Cai},
  \citenamefont {Babbush}, \citenamefont {Benjamin}, \citenamefont {Endo},
  \citenamefont {Huggins}, \citenamefont {Li}, \citenamefont {McClean},\ and\
  \citenamefont {O’Brien}}]{cai2023quantum}%
  \BibitemOpen
  \bibfield  {author} {\bibinfo {author} {\bibfnamefont {Z.}~\bibnamefont
  {Cai}}, \bibinfo {author} {\bibfnamefont {R.}~\bibnamefont {Babbush}},
  \bibinfo {author} {\bibfnamefont {S.~C.}\ \bibnamefont {Benjamin}}, \bibinfo
  {author} {\bibfnamefont {S.}~\bibnamefont {Endo}}, \bibinfo {author}
  {\bibfnamefont {W.~J.}\ \bibnamefont {Huggins}}, \bibinfo {author}
  {\bibfnamefont {Y.}~\bibnamefont {Li}}, \bibinfo {author} {\bibfnamefont
  {J.~R.}\ \bibnamefont {McClean}}, \ and\ \bibinfo {author} {\bibfnamefont
  {T.~E.}\ \bibnamefont {O’Brien}},\ }\bibfield  {title} {\enquote {\bibinfo
  {title} {Quantum error mitigation},}\ }\href@noop {} {\bibfield  {journal}
  {\bibinfo  {journal} {Reviews of Modern Physics}\ }\textbf {\bibinfo {volume}
  {95}},\ \bibinfo {pages} {045005} (\bibinfo {year} {2023})}\BibitemShut
  {NoStop}%
\bibitem [{\citenamefont {Li}\ and\ \citenamefont
  {Benjamin}(2017)}]{li2017efficient}%
  \BibitemOpen
  \bibfield  {author} {\bibinfo {author} {\bibfnamefont {Y.}~\bibnamefont
  {Li}}\ and\ \bibinfo {author} {\bibfnamefont {S.~C.}\ \bibnamefont
  {Benjamin}},\ }\bibfield  {title} {\enquote {\bibinfo {title} {Efficient
  variational quantum simulator incorporating active error minimization},}\
  }\href@noop {} {\bibfield  {journal} {\bibinfo  {journal} {Physical Review
  X}\ }\textbf {\bibinfo {volume} {7}},\ \bibinfo {pages} {021050} (\bibinfo
  {year} {2017})}\BibitemShut {NoStop}%
\bibitem [{\citenamefont {Temme}, \citenamefont {Bravyi},\ and\ \citenamefont
  {Gambetta}(2017)}]{Mitigation-PRL-2017}%
  \BibitemOpen
  \bibfield  {author} {\bibinfo {author} {\bibfnamefont {K.}~\bibnamefont
  {Temme}}, \bibinfo {author} {\bibfnamefont {S.}~\bibnamefont {Bravyi}}, \
  and\ \bibinfo {author} {\bibfnamefont {J.~M.}\ \bibnamefont {Gambetta}},\
  }\bibfield  {title} {\enquote {\bibinfo {title} {Error mitigation for
  short-depth quantum circuits},}\ }\href {\doibase
  10.1103/PhysRevLett.119.180509} {\bibfield  {journal} {\bibinfo  {journal}
  {Phys. Rev. Lett.}\ }\textbf {\bibinfo {volume} {119}},\ \bibinfo {pages}
  {180509} (\bibinfo {year} {2017})}\BibitemShut {NoStop}%
\bibitem [{\citenamefont {Giurgica-Tiron}\ \emph {et~al.}(2020)\citenamefont
  {Giurgica-Tiron}, \citenamefont {Hindy}, \citenamefont {LaRose},
  \citenamefont {Mari},\ and\ \citenamefont {Zeng}}]{giurgica2020digital}%
  \BibitemOpen
  \bibfield  {author} {\bibinfo {author} {\bibfnamefont {T.}~\bibnamefont
  {Giurgica-Tiron}}, \bibinfo {author} {\bibfnamefont {Y.}~\bibnamefont
  {Hindy}}, \bibinfo {author} {\bibfnamefont {R.}~\bibnamefont {LaRose}},
  \bibinfo {author} {\bibfnamefont {A.}~\bibnamefont {Mari}}, \ and\ \bibinfo
  {author} {\bibfnamefont {W.~J.}\ \bibnamefont {Zeng}},\ }\bibfield  {title}
  {\enquote {\bibinfo {title} {Digital zero noise extrapolation for quantum
  error mitigation},}\ }in\ \href@noop {} {\emph {\bibinfo {booktitle} {2020
  IEEE International Conference on Quantum Computing and Engineering (QCE)}}}\
  (\bibinfo {organization} {IEEE},\ \bibinfo {year} {2020})\ pp.\ \bibinfo
  {pages} {306--316}\BibitemShut {NoStop}%
\bibitem [{\citenamefont {Uvarov}\ \emph {et~al.}(2024)\citenamefont {Uvarov},
  \citenamefont {Rabinovich}, \citenamefont {Lakhmanskaya}, \citenamefont
  {Lakhmanskiy}, \citenamefont {Biamonte},\ and\ \citenamefont
  {Adhikary}}]{uvarov2024mitigating}%
  \BibitemOpen
  \bibfield  {author} {\bibinfo {author} {\bibfnamefont {A.}~\bibnamefont
  {Uvarov}}, \bibinfo {author} {\bibfnamefont {D.}~\bibnamefont {Rabinovich}},
  \bibinfo {author} {\bibfnamefont {O.}~\bibnamefont {Lakhmanskaya}}, \bibinfo
  {author} {\bibfnamefont {K.}~\bibnamefont {Lakhmanskiy}}, \bibinfo {author}
  {\bibfnamefont {J.}~\bibnamefont {Biamonte}}, \ and\ \bibinfo {author}
  {\bibfnamefont {S.}~\bibnamefont {Adhikary}},\ }\bibfield  {title} {\enquote
  {\bibinfo {title} {Mitigating quantum gate errors for variational
  eigensolvers using hardware-inspired zero-noise extrapolation},}\ }\href@noop
  {} {\bibfield  {journal} {\bibinfo  {journal} {Physical Review A}\ }\textbf
  {\bibinfo {volume} {110}},\ \bibinfo {pages} {012404} (\bibinfo {year}
  {2024})}\BibitemShut {NoStop}%
\bibitem [{\citenamefont {Garmon}, \citenamefont {Pooser},\ and\ \citenamefont
  {Dumitrescu}(2020)}]{garmon2020benchmarking}%
  \BibitemOpen
  \bibfield  {author} {\bibinfo {author} {\bibfnamefont {J.}~\bibnamefont
  {Garmon}}, \bibinfo {author} {\bibfnamefont {R.~C.}\ \bibnamefont {Pooser}},
  \ and\ \bibinfo {author} {\bibfnamefont {E.~F.}\ \bibnamefont {Dumitrescu}},\
  }\bibfield  {title} {\enquote {\bibinfo {title} {Benchmarking noise
  extrapolation with the openpulse control framework},}\ }\href@noop {}
  {\bibfield  {journal} {\bibinfo  {journal} {Physical Review A}\ }\textbf
  {\bibinfo {volume} {101}},\ \bibinfo {pages} {042308} (\bibinfo {year}
  {2020})}\BibitemShut {NoStop}%
\bibitem [{\citenamefont {Keen}\ \emph {et~al.}(2020)\citenamefont {Keen},
  \citenamefont {Maier}, \citenamefont {Johnston},\ and\ \citenamefont
  {Lougovski}}]{keen2020quantum}%
  \BibitemOpen
  \bibfield  {author} {\bibinfo {author} {\bibfnamefont {T.}~\bibnamefont
  {Keen}}, \bibinfo {author} {\bibfnamefont {T.}~\bibnamefont {Maier}},
  \bibinfo {author} {\bibfnamefont {S.}~\bibnamefont {Johnston}}, \ and\
  \bibinfo {author} {\bibfnamefont {P.}~\bibnamefont {Lougovski}},\ }\bibfield
  {title} {\enquote {\bibinfo {title} {Quantum-classical simulation of two-site
  dynamical mean-field theory on noisy quantum hardware},}\ }\href@noop {}
  {\bibfield  {journal} {\bibinfo  {journal} {Quantum Science and Technology}\
  }\textbf {\bibinfo {volume} {5}},\ \bibinfo {pages} {035001} (\bibinfo {year}
  {2020})}\BibitemShut {NoStop}%
\bibitem [{\citenamefont {Cai}(2021{\natexlab{a}})}]{mitigation_npj_2021}%
  \BibitemOpen
  \bibfield  {author} {\bibinfo {author} {\bibfnamefont {Z.}~\bibnamefont
  {Cai}},\ }\bibfield  {title} {\enquote {\bibinfo {title} {Multi-exponential
  error extrapolation and combining error mitigation techniques for nisq
  applications},}\ }\href {\doibase 10.1038/s41534-021-00404-3} {\bibfield
  {journal} {\bibinfo  {journal} {npj Quantum Information}\ }\textbf {\bibinfo
  {volume} {7}},\ \bibinfo {pages} {80} (\bibinfo {year}
  {2021}{\natexlab{a}})}\BibitemShut {NoStop}%
\bibitem [{\citenamefont {He}\ \emph {et~al.}(2020)\citenamefont {He},
  \citenamefont {Nachman}, \citenamefont {de~Jong},\ and\ \citenamefont
  {Bauer}}]{ZNE_PRA2020}%
  \BibitemOpen
  \bibfield  {author} {\bibinfo {author} {\bibfnamefont {A.}~\bibnamefont
  {He}}, \bibinfo {author} {\bibfnamefont {B.}~\bibnamefont {Nachman}},
  \bibinfo {author} {\bibfnamefont {W.~A.}\ \bibnamefont {de~Jong}}, \ and\
  \bibinfo {author} {\bibfnamefont {C.~W.}\ \bibnamefont {Bauer}},\ }\bibfield
  {title} {\enquote {\bibinfo {title} {Zero-noise extrapolation for
  quantum-gate error mitigation with identity insertions},}\ }\href {\doibase
  10.1103/PhysRevA.102.012426} {\bibfield  {journal} {\bibinfo  {journal}
  {Phys. Rev. A}\ }\textbf {\bibinfo {volume} {102}},\ \bibinfo {pages}
  {012426} (\bibinfo {year} {2020})}\BibitemShut {NoStop}%
\bibitem [{\citenamefont {Krebsbach}, \citenamefont {Trauzettel},\ and\
  \citenamefont {Calzona}(2022)}]{Richardson_extr_PRA2022}%
  \BibitemOpen
  \bibfield  {author} {\bibinfo {author} {\bibfnamefont {M.}~\bibnamefont
  {Krebsbach}}, \bibinfo {author} {\bibfnamefont {B.}~\bibnamefont
  {Trauzettel}}, \ and\ \bibinfo {author} {\bibfnamefont {A.}~\bibnamefont
  {Calzona}},\ }\bibfield  {title} {\enquote {\bibinfo {title} {Optimization of
  richardson extrapolation for quantum error mitigation},}\ }\href {\doibase
  10.1103/PhysRevA.106.062436} {\bibfield  {journal} {\bibinfo  {journal}
  {Phys. Rev. A}\ }\textbf {\bibinfo {volume} {106}},\ \bibinfo {pages}
  {062436} (\bibinfo {year} {2022})}\BibitemShut {NoStop}%
\bibitem [{\citenamefont {Russo}\ and\ \citenamefont
  {Mari}(2024)}]{Richardson_extr_PRA2024}%
  \BibitemOpen
  \bibfield  {author} {\bibinfo {author} {\bibfnamefont {V.}~\bibnamefont
  {Russo}}\ and\ \bibinfo {author} {\bibfnamefont {A.}~\bibnamefont {Mari}},\
  }\bibfield  {title} {\enquote {\bibinfo {title} {Quantum error mitigation by
  layerwise richardson extrapolation},}\ }\href {\doibase
  10.1103/PhysRevA.110.062420} {\bibfield  {journal} {\bibinfo  {journal}
  {Phys. Rev. A}\ }\textbf {\bibinfo {volume} {110}},\ \bibinfo {pages}
  {062420} (\bibinfo {year} {2024})}\BibitemShut {NoStop}%
\bibitem [{\citenamefont {Cervera-Lierta}(2018)}]{CerveraIsing2018}%
  \BibitemOpen
  \bibfield  {author} {\bibinfo {author} {\bibfnamefont {A.}~\bibnamefont
  {Cervera-Lierta}},\ }\bibfield  {title} {\enquote {\bibinfo {title} {Exact
  {I}sing model simulation on a quantum computer},}\ }\href {\doibase
  10.22331/q-2018-12-21-114} {\bibfield  {journal} {\bibinfo  {journal}
  {{Quantum}}\ }\textbf {\bibinfo {volume} {2}},\ \bibinfo {pages} {114}
  (\bibinfo {year} {2018})}\BibitemShut {NoStop}%
\bibitem [{\citenamefont {Mi}\ \emph {et~al.}(2022)\citenamefont {Mi},
  \citenamefont {Ippoliti}, \citenamefont {Quintana}, \citenamefont {Greene},
  \citenamefont {Chen}, \citenamefont {Gross}, \citenamefont {Arute},
  \citenamefont {Arya}, \citenamefont {Atalaya}, \citenamefont {Babbush} \emph
  {et~al.}}]{mi2022time}%
  \BibitemOpen
  \bibfield  {author} {\bibinfo {author} {\bibfnamefont {X.}~\bibnamefont
  {Mi}}, \bibinfo {author} {\bibfnamefont {M.}~\bibnamefont {Ippoliti}},
  \bibinfo {author} {\bibfnamefont {C.}~\bibnamefont {Quintana}}, \bibinfo
  {author} {\bibfnamefont {A.}~\bibnamefont {Greene}}, \bibinfo {author}
  {\bibfnamefont {Z.}~\bibnamefont {Chen}}, \bibinfo {author} {\bibfnamefont
  {J.}~\bibnamefont {Gross}}, \bibinfo {author} {\bibfnamefont
  {F.}~\bibnamefont {Arute}}, \bibinfo {author} {\bibfnamefont
  {K.}~\bibnamefont {Arya}}, \bibinfo {author} {\bibfnamefont {J.}~\bibnamefont
  {Atalaya}}, \bibinfo {author} {\bibfnamefont {R.}~\bibnamefont {Babbush}},
  \emph {et~al.},\ }\bibfield  {title} {\enquote {\bibinfo {title}
  {Time-crystalline eigenstate order on a quantum processor},}\ }\href@noop {}
  {\bibfield  {journal} {\bibinfo  {journal} {Nature}\ }\textbf {\bibinfo
  {volume} {601}},\ \bibinfo {pages} {531--536} (\bibinfo {year}
  {2022})}\BibitemShut {NoStop}%
\bibitem [{\citenamefont {Frey}\ and\ \citenamefont
  {Rachel}(2022)}]{frey2022realization}%
  \BibitemOpen
  \bibfield  {author} {\bibinfo {author} {\bibfnamefont {P.}~\bibnamefont
  {Frey}}\ and\ \bibinfo {author} {\bibfnamefont {S.}~\bibnamefont {Rachel}},\
  }\bibfield  {title} {\enquote {\bibinfo {title} {Realization of a discrete
  time crystal on 57 qubits of a quantum computer},}\ }\href@noop {} {\bibfield
   {journal} {\bibinfo  {journal} {Science advances}\ }\textbf {\bibinfo
  {volume} {8}},\ \bibinfo {pages} {eabm7652} (\bibinfo {year}
  {2022})}\BibitemShut {NoStop}%
\bibitem [{\citenamefont {Chen}\ \emph {et~al.}(2022)\citenamefont {Chen},
  \citenamefont {Burdick}, \citenamefont {Yao}, \citenamefont {Orth},\ and\
  \citenamefont {Iadecola}}]{chen2022error}%
  \BibitemOpen
  \bibfield  {author} {\bibinfo {author} {\bibfnamefont {I.-C.}\ \bibnamefont
  {Chen}}, \bibinfo {author} {\bibfnamefont {B.}~\bibnamefont {Burdick}},
  \bibinfo {author} {\bibfnamefont {Y.}~\bibnamefont {Yao}}, \bibinfo {author}
  {\bibfnamefont {P.~P.}\ \bibnamefont {Orth}}, \ and\ \bibinfo {author}
  {\bibfnamefont {T.}~\bibnamefont {Iadecola}},\ }\bibfield  {title} {\enquote
  {\bibinfo {title} {Error-mitigated simulation of quantum many-body scars on
  quantum computers with pulse-level control},}\ }\href@noop {} {\bibfield
  {journal} {\bibinfo  {journal} {Physical Review Research}\ }\textbf {\bibinfo
  {volume} {4}},\ \bibinfo {pages} {043027} (\bibinfo {year}
  {2022})}\BibitemShut {NoStop}%
\bibitem [{\citenamefont {Rigetti}\ \emph {et~al.}(2012)\citenamefont
  {Rigetti}, \citenamefont {Gambetta}, \citenamefont {Poletto}, \citenamefont
  {Plourde}, \citenamefont {Chow}, \citenamefont {C{\'o}rcoles}, \citenamefont
  {Smolin}, \citenamefont {Merkel}, \citenamefont {Rozen}, \citenamefont
  {Keefe} \emph {et~al.}}]{rigetti2012superconducting}%
  \BibitemOpen
  \bibfield  {author} {\bibinfo {author} {\bibfnamefont {C.}~\bibnamefont
  {Rigetti}}, \bibinfo {author} {\bibfnamefont {J.~M.}\ \bibnamefont
  {Gambetta}}, \bibinfo {author} {\bibfnamefont {S.}~\bibnamefont {Poletto}},
  \bibinfo {author} {\bibfnamefont {B.~L.}\ \bibnamefont {Plourde}}, \bibinfo
  {author} {\bibfnamefont {J.~M.}\ \bibnamefont {Chow}}, \bibinfo {author}
  {\bibfnamefont {A.~D.}\ \bibnamefont {C{\'o}rcoles}}, \bibinfo {author}
  {\bibfnamefont {J.~A.}\ \bibnamefont {Smolin}}, \bibinfo {author}
  {\bibfnamefont {S.~T.}\ \bibnamefont {Merkel}}, \bibinfo {author}
  {\bibfnamefont {J.~R.}\ \bibnamefont {Rozen}}, \bibinfo {author}
  {\bibfnamefont {G.~A.}\ \bibnamefont {Keefe}},  \emph {et~al.},\ }\bibfield
  {title} {\enquote {\bibinfo {title} {Superconducting qubit in a waveguide
  cavity with a coherence time approaching 0.1 ms},}\ }\href@noop {} {\bibfield
   {journal} {\bibinfo  {journal} {Physical Review B—Condensed Matter and
  Materials Physics}\ }\textbf {\bibinfo {volume} {86}},\ \bibinfo {pages}
  {100506} (\bibinfo {year} {2012})}\BibitemShut {NoStop}%
\bibitem [{\citenamefont {Place}\ \emph {et~al.}(2021)\citenamefont {Place},
  \citenamefont {Rodgers}, \citenamefont {Mundada}, \citenamefont {Smitham},
  \citenamefont {Fitzpatrick}, \citenamefont {Leng}, \citenamefont {Premkumar},
  \citenamefont {Bryon}, \citenamefont {Vrajitoarea}, \citenamefont {Sussman}
  \emph {et~al.}}]{place2021new}%
  \BibitemOpen
  \bibfield  {author} {\bibinfo {author} {\bibfnamefont {A.~P.}\ \bibnamefont
  {Place}}, \bibinfo {author} {\bibfnamefont {L.~V.}\ \bibnamefont {Rodgers}},
  \bibinfo {author} {\bibfnamefont {P.}~\bibnamefont {Mundada}}, \bibinfo
  {author} {\bibfnamefont {B.~M.}\ \bibnamefont {Smitham}}, \bibinfo {author}
  {\bibfnamefont {M.}~\bibnamefont {Fitzpatrick}}, \bibinfo {author}
  {\bibfnamefont {Z.}~\bibnamefont {Leng}}, \bibinfo {author} {\bibfnamefont
  {A.}~\bibnamefont {Premkumar}}, \bibinfo {author} {\bibfnamefont
  {J.}~\bibnamefont {Bryon}}, \bibinfo {author} {\bibfnamefont
  {A.}~\bibnamefont {Vrajitoarea}}, \bibinfo {author} {\bibfnamefont
  {S.}~\bibnamefont {Sussman}},  \emph {et~al.},\ }\bibfield  {title} {\enquote
  {\bibinfo {title} {New material platform for superconducting transmon qubits
  with coherence times exceeding 0.3 milliseconds},}\ }\href@noop {} {\bibfield
   {journal} {\bibinfo  {journal} {Nature communications}\ }\textbf {\bibinfo
  {volume} {12}},\ \bibinfo {pages} {1779} (\bibinfo {year}
  {2021})}\BibitemShut {NoStop}%
\bibitem [{\citenamefont {Sheldon}\ \emph {et~al.}(2016)\citenamefont
  {Sheldon}, \citenamefont {Magesan}, \citenamefont {Chow},\ and\ \citenamefont
  {Gambetta}}]{sheldon2016procedure}%
  \BibitemOpen
  \bibfield  {author} {\bibinfo {author} {\bibfnamefont {S.}~\bibnamefont
  {Sheldon}}, \bibinfo {author} {\bibfnamefont {E.}~\bibnamefont {Magesan}},
  \bibinfo {author} {\bibfnamefont {J.~M.}\ \bibnamefont {Chow}}, \ and\
  \bibinfo {author} {\bibfnamefont {J.~M.}\ \bibnamefont {Gambetta}},\
  }\bibfield  {title} {\enquote {\bibinfo {title} {Procedure for systematically
  tuning up cross-talk in the cross-resonance gate},}\ }\href@noop {}
  {\bibfield  {journal} {\bibinfo  {journal} {Physical Review A}\ }\textbf
  {\bibinfo {volume} {93}},\ \bibinfo {pages} {060302} (\bibinfo {year}
  {2016})}\BibitemShut {NoStop}%
\bibitem [{\citenamefont {Dial}(2022)}]{dial2022eagle}%
  \BibitemOpen
  \bibfield  {author} {\bibinfo {author} {\bibfnamefont {O.}~\bibnamefont
  {Dial}},\ }\href@noop {} {\enquote {\bibinfo {title} {Eagle’s quantum
  performance progress},}\ } (\bibinfo {year} {2022})\BibitemShut {NoStop}%
\bibitem [{\citenamefont {Chow}, \citenamefont {Dial},\ and\ \citenamefont
  {Gambetta}(2021)}]{chow2021ibm}%
  \BibitemOpen
  \bibfield  {author} {\bibinfo {author} {\bibfnamefont {J.}~\bibnamefont
  {Chow}}, \bibinfo {author} {\bibfnamefont {O.}~\bibnamefont {Dial}}, \ and\
  \bibinfo {author} {\bibfnamefont {J.}~\bibnamefont {Gambetta}},\ }\bibfield
  {title} {\enquote {\bibinfo {title} {Ibm quantum breaks the 100-qubit
  processor barrier},}\ }\href@noop {} {\bibfield  {journal} {\bibinfo
  {journal} {IBM Research Blog}\ }\textbf {\bibinfo {volume} {2}} (\bibinfo
  {year} {2021})}\BibitemShut {NoStop}%
\bibitem [{\citenamefont {Aleksandrowicz}\ \emph {et~al.}(2019)\citenamefont
  {Aleksandrowicz}, \citenamefont {Alexander}, \citenamefont {Barkoutsos},
  \citenamefont {Bello}, \citenamefont {Ben-Haim}, \citenamefont {Bucher},
  \citenamefont {Cabrera-Hernández}, \citenamefont {Carballo-Franquis},
  \citenamefont {Chen}, \citenamefont {Chen}, \citenamefont {Chow},
  \citenamefont {Córcoles-Gonzales}, \citenamefont {Cross}, \citenamefont
  {Cross}, \citenamefont {Cruz-Benito}, \citenamefont {Culver}, \citenamefont
  {González}, \citenamefont {Torre}, \citenamefont {Ding}, \citenamefont
  {Dumitrescu}, \citenamefont {Duran}, \citenamefont {Eendebak}, \citenamefont
  {Everitt}, \citenamefont {Sertage}, \citenamefont {Frisch}, \citenamefont
  {Fuhrer}, \citenamefont {Gambetta}, \citenamefont {Gago}, \citenamefont
  {Gomez-Mosquera}, \citenamefont {Greenberg}, \citenamefont {Hamamura},
  \citenamefont {Havlicek}, \citenamefont {Hellmers}, \citenamefont {Łukasz
  Herok}, \citenamefont {Horii}, \citenamefont {Hu}, \citenamefont {Imamichi},
  \citenamefont {Itoko}, \citenamefont {Javadi-Abhari}, \citenamefont
  {Kanazawa}, \citenamefont {Karazeev}, \citenamefont {Krsulich}, \citenamefont
  {Liu}, \citenamefont {Luh}, \citenamefont {Maeng}, \citenamefont {Marques},
  \citenamefont {Martín-Fernández}, \citenamefont {McClure}, \citenamefont
  {McKay}, \citenamefont {Meesala}, \citenamefont {Mezzacapo}, \citenamefont
  {Moll}, \citenamefont {Rodríguez}, \citenamefont {Nannicini}, \citenamefont
  {Nation}, \citenamefont {Ollitrault}, \citenamefont {O'Riordan},
  \citenamefont {Paik}, \citenamefont {Pérez}, \citenamefont {Phan},
  \citenamefont {Pistoia}, \citenamefont {Prutyanov}, \citenamefont {Reuter},
  \citenamefont {Rice}, \citenamefont {Davila}, \citenamefont {Rudy},
  \citenamefont {Ryu}, \citenamefont {Sathaye}, \citenamefont {Schnabel},
  \citenamefont {Schoute}, \citenamefont {Setia}, \citenamefont {Shi},
  \citenamefont {Silva}, \citenamefont {Siraichi}, \citenamefont {Sivarajah},
  \citenamefont {Smolin}, \citenamefont {Soeken}, \citenamefont {Takahashi},
  \citenamefont {Tavernelli}, \citenamefont {Taylor}, \citenamefont {Taylour},
  \citenamefont {Trabing}, \citenamefont {Treinish}, \citenamefont {Turner},
  \citenamefont {Vogt-Lee}, \citenamefont {Vuillot}, \citenamefont {Wildstrom},
  \citenamefont {Wilson}, \citenamefont {Winston}, \citenamefont {Wood},
  \citenamefont {Wood}, \citenamefont {Wörner}, \citenamefont {Akhalwaya},\
  and\ \citenamefont {Zoufal}}]{abraham2019qiskit}%
  \BibitemOpen
  \bibfield  {author} {\bibinfo {author} {\bibfnamefont {G.}~\bibnamefont
  {Aleksandrowicz}}, \bibinfo {author} {\bibfnamefont {T.}~\bibnamefont
  {Alexander}}, \bibinfo {author} {\bibfnamefont {P.}~\bibnamefont
  {Barkoutsos}}, \bibinfo {author} {\bibfnamefont {L.}~\bibnamefont {Bello}},
  \bibinfo {author} {\bibfnamefont {Y.}~\bibnamefont {Ben-Haim}}, \bibinfo
  {author} {\bibfnamefont {D.}~\bibnamefont {Bucher}}, \bibinfo {author}
  {\bibfnamefont {F.~J.}\ \bibnamefont {Cabrera-Hernández}}, \bibinfo {author}
  {\bibfnamefont {J.}~\bibnamefont {Carballo-Franquis}}, \bibinfo {author}
  {\bibfnamefont {A.}~\bibnamefont {Chen}}, \bibinfo {author} {\bibfnamefont
  {C.-F.}\ \bibnamefont {Chen}}, \bibinfo {author} {\bibfnamefont {J.~M.}\
  \bibnamefont {Chow}}, \bibinfo {author} {\bibfnamefont {A.~D.}\ \bibnamefont
  {Córcoles-Gonzales}}, \bibinfo {author} {\bibfnamefont {A.~J.}\ \bibnamefont
  {Cross}}, \bibinfo {author} {\bibfnamefont {A.}~\bibnamefont {Cross}},
  \bibinfo {author} {\bibfnamefont {J.}~\bibnamefont {Cruz-Benito}}, \bibinfo
  {author} {\bibfnamefont {C.}~\bibnamefont {Culver}}, \bibinfo {author}
  {\bibfnamefont {S.~D. L.~P.}\ \bibnamefont {González}}, \bibinfo {author}
  {\bibfnamefont {E.~D.~L.}\ \bibnamefont {Torre}}, \bibinfo {author}
  {\bibfnamefont {D.}~\bibnamefont {Ding}}, \bibinfo {author} {\bibfnamefont
  {E.}~\bibnamefont {Dumitrescu}}, \bibinfo {author} {\bibfnamefont
  {I.}~\bibnamefont {Duran}}, \bibinfo {author} {\bibfnamefont
  {P.}~\bibnamefont {Eendebak}}, \bibinfo {author} {\bibfnamefont
  {M.}~\bibnamefont {Everitt}}, \bibinfo {author} {\bibfnamefont {I.~F.}\
  \bibnamefont {Sertage}}, \bibinfo {author} {\bibfnamefont {A.}~\bibnamefont
  {Frisch}}, \bibinfo {author} {\bibfnamefont {A.}~\bibnamefont {Fuhrer}},
  \bibinfo {author} {\bibfnamefont {J.}~\bibnamefont {Gambetta}}, \bibinfo
  {author} {\bibfnamefont {B.~G.}\ \bibnamefont {Gago}}, \bibinfo {author}
  {\bibfnamefont {J.}~\bibnamefont {Gomez-Mosquera}}, \bibinfo {author}
  {\bibfnamefont {D.}~\bibnamefont {Greenberg}}, \bibinfo {author}
  {\bibfnamefont {I.}~\bibnamefont {Hamamura}}, \bibinfo {author}
  {\bibfnamefont {V.}~\bibnamefont {Havlicek}}, \bibinfo {author}
  {\bibfnamefont {J.}~\bibnamefont {Hellmers}}, \bibinfo {author} {\bibnamefont
  {Łukasz Herok}}, \bibinfo {author} {\bibfnamefont {H.}~\bibnamefont
  {Horii}}, \bibinfo {author} {\bibfnamefont {S.}~\bibnamefont {Hu}}, \bibinfo
  {author} {\bibfnamefont {T.}~\bibnamefont {Imamichi}}, \bibinfo {author}
  {\bibfnamefont {T.}~\bibnamefont {Itoko}}, \bibinfo {author} {\bibfnamefont
  {A.}~\bibnamefont {Javadi-Abhari}}, \bibinfo {author} {\bibfnamefont
  {N.}~\bibnamefont {Kanazawa}}, \bibinfo {author} {\bibfnamefont
  {A.}~\bibnamefont {Karazeev}}, \bibinfo {author} {\bibfnamefont
  {K.}~\bibnamefont {Krsulich}}, \bibinfo {author} {\bibfnamefont
  {P.}~\bibnamefont {Liu}}, \bibinfo {author} {\bibfnamefont {Y.}~\bibnamefont
  {Luh}}, \bibinfo {author} {\bibfnamefont {Y.}~\bibnamefont {Maeng}}, \bibinfo
  {author} {\bibfnamefont {M.}~\bibnamefont {Marques}}, \bibinfo {author}
  {\bibfnamefont {F.~J.}\ \bibnamefont {Martín-Fernández}}, \bibinfo {author}
  {\bibfnamefont {D.~T.}\ \bibnamefont {McClure}}, \bibinfo {author}
  {\bibfnamefont {D.}~\bibnamefont {McKay}}, \bibinfo {author} {\bibfnamefont
  {S.}~\bibnamefont {Meesala}}, \bibinfo {author} {\bibfnamefont
  {A.}~\bibnamefont {Mezzacapo}}, \bibinfo {author} {\bibfnamefont
  {N.}~\bibnamefont {Moll}}, \bibinfo {author} {\bibfnamefont {D.~M.}\
  \bibnamefont {Rodríguez}}, \bibinfo {author} {\bibfnamefont
  {G.}~\bibnamefont {Nannicini}}, \bibinfo {author} {\bibfnamefont
  {P.}~\bibnamefont {Nation}}, \bibinfo {author} {\bibfnamefont
  {P.}~\bibnamefont {Ollitrault}}, \bibinfo {author} {\bibfnamefont {L.~J.}\
  \bibnamefont {O'Riordan}}, \bibinfo {author} {\bibfnamefont {H.}~\bibnamefont
  {Paik}}, \bibinfo {author} {\bibfnamefont {J.}~\bibnamefont {Pérez}},
  \bibinfo {author} {\bibfnamefont {A.}~\bibnamefont {Phan}}, \bibinfo {author}
  {\bibfnamefont {M.}~\bibnamefont {Pistoia}}, \bibinfo {author} {\bibfnamefont
  {V.}~\bibnamefont {Prutyanov}}, \bibinfo {author} {\bibfnamefont
  {M.}~\bibnamefont {Reuter}}, \bibinfo {author} {\bibfnamefont
  {J.}~\bibnamefont {Rice}}, \bibinfo {author} {\bibfnamefont {A.~R.}\
  \bibnamefont {Davila}}, \bibinfo {author} {\bibfnamefont {R.~H.~P.}\
  \bibnamefont {Rudy}}, \bibinfo {author} {\bibfnamefont {M.}~\bibnamefont
  {Ryu}}, \bibinfo {author} {\bibfnamefont {N.}~\bibnamefont {Sathaye}},
  \bibinfo {author} {\bibfnamefont {C.}~\bibnamefont {Schnabel}}, \bibinfo
  {author} {\bibfnamefont {E.}~\bibnamefont {Schoute}}, \bibinfo {author}
  {\bibfnamefont {K.}~\bibnamefont {Setia}}, \bibinfo {author} {\bibfnamefont
  {Y.}~\bibnamefont {Shi}}, \bibinfo {author} {\bibfnamefont {A.}~\bibnamefont
  {Silva}}, \bibinfo {author} {\bibfnamefont {Y.}~\bibnamefont {Siraichi}},
  \bibinfo {author} {\bibfnamefont {S.}~\bibnamefont {Sivarajah}}, \bibinfo
  {author} {\bibfnamefont {J.~A.}\ \bibnamefont {Smolin}}, \bibinfo {author}
  {\bibfnamefont {M.}~\bibnamefont {Soeken}}, \bibinfo {author} {\bibfnamefont
  {H.}~\bibnamefont {Takahashi}}, \bibinfo {author} {\bibfnamefont
  {I.}~\bibnamefont {Tavernelli}}, \bibinfo {author} {\bibfnamefont
  {C.}~\bibnamefont {Taylor}}, \bibinfo {author} {\bibfnamefont
  {P.}~\bibnamefont {Taylour}}, \bibinfo {author} {\bibfnamefont
  {K.}~\bibnamefont {Trabing}}, \bibinfo {author} {\bibfnamefont
  {M.}~\bibnamefont {Treinish}}, \bibinfo {author} {\bibfnamefont
  {W.}~\bibnamefont {Turner}}, \bibinfo {author} {\bibfnamefont
  {D.}~\bibnamefont {Vogt-Lee}}, \bibinfo {author} {\bibfnamefont
  {C.}~\bibnamefont {Vuillot}}, \bibinfo {author} {\bibfnamefont {J.~A.}\
  \bibnamefont {Wildstrom}}, \bibinfo {author} {\bibfnamefont {J.}~\bibnamefont
  {Wilson}}, \bibinfo {author} {\bibfnamefont {E.}~\bibnamefont {Winston}},
  \bibinfo {author} {\bibfnamefont {C.}~\bibnamefont {Wood}}, \bibinfo {author}
  {\bibfnamefont {S.}~\bibnamefont {Wood}}, \bibinfo {author} {\bibfnamefont
  {S.}~\bibnamefont {Wörner}}, \bibinfo {author} {\bibfnamefont {I.~Y.}\
  \bibnamefont {Akhalwaya}}, \ and\ \bibinfo {author} {\bibfnamefont
  {C.}~\bibnamefont {Zoufal}},\ }\href {\doibase 10.5281/zenodo.2562111}
  {\enquote {\bibinfo {title} {{Qiskit: An Open-source Framework for Quantum
  Computing}},}\ } (\bibinfo {year} {2019})\BibitemShut {NoStop}%
\bibitem [{\citenamefont {Van Den~Berg}, \citenamefont {Minev},\ and\
  \citenamefont {Temme}(2022)}]{van2022model}%
  \BibitemOpen
  \bibfield  {author} {\bibinfo {author} {\bibfnamefont {E.}~\bibnamefont {Van
  Den~Berg}}, \bibinfo {author} {\bibfnamefont {Z.~K.}\ \bibnamefont {Minev}},
  \ and\ \bibinfo {author} {\bibfnamefont {K.}~\bibnamefont {Temme}},\
  }\bibfield  {title} {\enquote {\bibinfo {title} {Model-free readout-error
  mitigation for quantum expectation values},}\ }\href@noop {} {\bibfield
  {journal} {\bibinfo  {journal} {Physical Review A}\ }\textbf {\bibinfo
  {volume} {105}},\ \bibinfo {pages} {032620} (\bibinfo {year}
  {2022})}\BibitemShut {NoStop}%
\bibitem [{\citenamefont {Chen}\ \emph {et~al.}(2019)\citenamefont {Chen},
  \citenamefont {Farahzad}, \citenamefont {Yoo},\ and\ \citenamefont
  {Wei}}]{chen2019detector}%
  \BibitemOpen
  \bibfield  {author} {\bibinfo {author} {\bibfnamefont {Y.}~\bibnamefont
  {Chen}}, \bibinfo {author} {\bibfnamefont {M.}~\bibnamefont {Farahzad}},
  \bibinfo {author} {\bibfnamefont {S.}~\bibnamefont {Yoo}}, \ and\ \bibinfo
  {author} {\bibfnamefont {T.-C.}\ \bibnamefont {Wei}},\ }\bibfield  {title}
  {\enquote {\bibinfo {title} {Detector tomography on ibm quantum computers and
  mitigation of an imperfect measurement},}\ }\href@noop {} {\bibfield
  {journal} {\bibinfo  {journal} {Physical Review A}\ }\textbf {\bibinfo
  {volume} {100}},\ \bibinfo {pages} {052315} (\bibinfo {year}
  {2019})}\BibitemShut {NoStop}%
\bibitem [{\citenamefont {Bravyi}\ \emph {et~al.}(2021)\citenamefont {Bravyi},
  \citenamefont {Sheldon}, \citenamefont {Kandala}, \citenamefont {Mckay},\
  and\ \citenamefont {Gambetta}}]{bravyi2021mitigating}%
  \BibitemOpen
  \bibfield  {author} {\bibinfo {author} {\bibfnamefont {S.}~\bibnamefont
  {Bravyi}}, \bibinfo {author} {\bibfnamefont {S.}~\bibnamefont {Sheldon}},
  \bibinfo {author} {\bibfnamefont {A.}~\bibnamefont {Kandala}}, \bibinfo
  {author} {\bibfnamefont {D.~C.}\ \bibnamefont {Mckay}}, \ and\ \bibinfo
  {author} {\bibfnamefont {J.~M.}\ \bibnamefont {Gambetta}},\ }\bibfield
  {title} {\enquote {\bibinfo {title} {Mitigating measurement errors in
  multiqubit experiments},}\ }\href@noop {} {\bibfield  {journal} {\bibinfo
  {journal} {Physical Review A}\ }\textbf {\bibinfo {volume} {103}},\ \bibinfo
  {pages} {042605} (\bibinfo {year} {2021})}\BibitemShut {NoStop}%
\bibitem [{\citenamefont {Wallman}\ and\ \citenamefont
  {Emerson}(2016)}]{Twirling2016}%
  \BibitemOpen
  \bibfield  {author} {\bibinfo {author} {\bibfnamefont {J.~J.}\ \bibnamefont
  {Wallman}}\ and\ \bibinfo {author} {\bibfnamefont {J.}~\bibnamefont
  {Emerson}},\ }\bibfield  {title} {\enquote {\bibinfo {title} {Noise tailoring
  for scalable quantum computation via randomized compiling},}\ }\href
  {\doibase 10.1103/PhysRevA.94.052325} {\bibfield  {journal} {\bibinfo
  {journal} {Phys. Rev. A}\ }\textbf {\bibinfo {volume} {94}},\ \bibinfo
  {pages} {052325} (\bibinfo {year} {2016})}\BibitemShut {NoStop}%
\bibitem [{not()}]{note_Groover_prob}%
  \BibitemOpen
  \href@noop {} {}\bibinfo {note} {In a noisless evaluation of the Grover's
  algorithm, the probability of finding the target state after k iterations is
  $P = \sin^2\left((2k+1)\theta\right)$ with $\theta=\arcsin(2^{-n/2})$. Since
  we consider $k\approx\frac{\pi}{4}\sqrt{2^{n}}-\frac{1}{2}$ we can write $P =
  \sin^2\left(\left(\pi2^{\frac{n}{2}-1}+1\right)\arcsin\left(2^{-\frac{n}{2}}\right)\right)$}\BibitemShut
  {NoStop}%
\bibitem [{\citenamefont {Watrous}(2006)}]{WatrousQCNotes}%
  \BibitemOpen
  \bibfield  {author} {\bibinfo {author} {\bibfnamefont {J.}~\bibnamefont
  {Watrous}},\ }\href@noop {} {\enquote {\bibinfo {title} {Lecture notes on
  quantum computation},}\ }\bibinfo {howpublished}
  {\url{https://cs.uwaterloo.ca/~watrous/QC-notes/QC-notes.13.pdf}} (\bibinfo
  {year} {2006})\BibitemShut {NoStop}%
\bibitem [{\citenamefont {Verstraete}, \citenamefont {Cirac},\ and\
  \citenamefont {Latorre}(2009)}]{LaTorrePRA2009}%
  \BibitemOpen
  \bibfield  {author} {\bibinfo {author} {\bibfnamefont {F.}~\bibnamefont
  {Verstraete}}, \bibinfo {author} {\bibfnamefont {J.~I.}\ \bibnamefont
  {Cirac}}, \ and\ \bibinfo {author} {\bibfnamefont {J.~I.}\ \bibnamefont
  {Latorre}},\ }\bibfield  {title} {\enquote {\bibinfo {title} {Quantum
  circuits for strongly correlated quantum systems},}\ }\href {\doibase
  10.1103/PhysRevA.79.032316} {\bibfield  {journal} {\bibinfo  {journal} {Phys.
  Rev. A}\ }\textbf {\bibinfo {volume} {79}},\ \bibinfo {pages} {032316}
  (\bibinfo {year} {2009})}\BibitemShut {NoStop}%
\bibitem [{\citenamefont {Arrazola}\ \emph {et~al.}(2021)\citenamefont
  {Arrazola}, \citenamefont {Bergholm}, \citenamefont {Br{\'a}dler},
  \citenamefont {Bromley}, \citenamefont {Collins}, \citenamefont {Dhand},
  \citenamefont {Fumagalli}, \citenamefont {Gerrits}, \citenamefont {Goussev},
  \citenamefont {Helt}, \citenamefont {Hundal}, \citenamefont {Isacsson},
  \citenamefont {Israel}, \citenamefont {Izaac}, \citenamefont {Jahangiri},
  \citenamefont {Janik}, \citenamefont {Killoran}, \citenamefont {Kumar},
  \citenamefont {Lavoie}, \citenamefont {Lita}, \citenamefont {Mahler},
  \citenamefont {Menotti}, \citenamefont {Morrison}, \citenamefont {Nam},
  \citenamefont {Neuhaus}, \citenamefont {Qi}, \citenamefont {Quesada},
  \citenamefont {Repingon}, \citenamefont {Sabapathy}, \citenamefont {Schuld},
  \citenamefont {Su}, \citenamefont {Swinarton}, \citenamefont {Sz{\'a}va},
  \citenamefont {Tan}, \citenamefont {Tan}, \citenamefont {Vaidya},
  \citenamefont {Vernon}, \citenamefont {Zabaneh},\ and\ \citenamefont
  {Zhang}}]{Arrazola2021}%
  \BibitemOpen
  \bibfield  {author} {\bibinfo {author} {\bibfnamefont {J.~M.}\ \bibnamefont
  {Arrazola}}, \bibinfo {author} {\bibfnamefont {V.}~\bibnamefont {Bergholm}},
  \bibinfo {author} {\bibfnamefont {K.}~\bibnamefont {Br{\'a}dler}}, \bibinfo
  {author} {\bibfnamefont {T.~R.}\ \bibnamefont {Bromley}}, \bibinfo {author}
  {\bibfnamefont {M.~J.}\ \bibnamefont {Collins}}, \bibinfo {author}
  {\bibfnamefont {I.}~\bibnamefont {Dhand}}, \bibinfo {author} {\bibfnamefont
  {A.}~\bibnamefont {Fumagalli}}, \bibinfo {author} {\bibfnamefont
  {T.}~\bibnamefont {Gerrits}}, \bibinfo {author} {\bibfnamefont
  {A.}~\bibnamefont {Goussev}}, \bibinfo {author} {\bibfnamefont {L.~G.}\
  \bibnamefont {Helt}}, \bibinfo {author} {\bibfnamefont {J.}~\bibnamefont
  {Hundal}}, \bibinfo {author} {\bibfnamefont {T.}~\bibnamefont {Isacsson}},
  \bibinfo {author} {\bibfnamefont {R.~B.}\ \bibnamefont {Israel}}, \bibinfo
  {author} {\bibfnamefont {J.}~\bibnamefont {Izaac}}, \bibinfo {author}
  {\bibfnamefont {S.}~\bibnamefont {Jahangiri}}, \bibinfo {author}
  {\bibfnamefont {R.}~\bibnamefont {Janik}}, \bibinfo {author} {\bibfnamefont
  {N.}~\bibnamefont {Killoran}}, \bibinfo {author} {\bibfnamefont {S.~P.}\
  \bibnamefont {Kumar}}, \bibinfo {author} {\bibfnamefont {J.}~\bibnamefont
  {Lavoie}}, \bibinfo {author} {\bibfnamefont {A.~E.}\ \bibnamefont {Lita}},
  \bibinfo {author} {\bibfnamefont {D.~H.}\ \bibnamefont {Mahler}}, \bibinfo
  {author} {\bibfnamefont {M.}~\bibnamefont {Menotti}}, \bibinfo {author}
  {\bibfnamefont {B.}~\bibnamefont {Morrison}}, \bibinfo {author}
  {\bibfnamefont {S.~W.}\ \bibnamefont {Nam}}, \bibinfo {author} {\bibfnamefont
  {L.}~\bibnamefont {Neuhaus}}, \bibinfo {author} {\bibfnamefont {H.~Y.}\
  \bibnamefont {Qi}}, \bibinfo {author} {\bibfnamefont {N.}~\bibnamefont
  {Quesada}}, \bibinfo {author} {\bibfnamefont {A.}~\bibnamefont {Repingon}},
  \bibinfo {author} {\bibfnamefont {K.~K.}\ \bibnamefont {Sabapathy}}, \bibinfo
  {author} {\bibfnamefont {M.}~\bibnamefont {Schuld}}, \bibinfo {author}
  {\bibfnamefont {D.}~\bibnamefont {Su}}, \bibinfo {author} {\bibfnamefont
  {J.}~\bibnamefont {Swinarton}}, \bibinfo {author} {\bibfnamefont
  {A.}~\bibnamefont {Sz{\'a}va}}, \bibinfo {author} {\bibfnamefont
  {K.}~\bibnamefont {Tan}}, \bibinfo {author} {\bibfnamefont {P.}~\bibnamefont
  {Tan}}, \bibinfo {author} {\bibfnamefont {V.~D.}\ \bibnamefont {Vaidya}},
  \bibinfo {author} {\bibfnamefont {Z.}~\bibnamefont {Vernon}}, \bibinfo
  {author} {\bibfnamefont {Z.}~\bibnamefont {Zabaneh}}, \ and\ \bibinfo
  {author} {\bibfnamefont {Y.}~\bibnamefont {Zhang}},\ }\bibfield  {title}
  {\enquote {\bibinfo {title} {Quantum circuits with many photons on a
  programmable nanophotonic chip},}\ }\href {\doibase
  10.1038/s41586-021-03202-1} {\bibfield  {journal} {\bibinfo  {journal}
  {Nature}\ }\textbf {\bibinfo {volume} {591}},\ \bibinfo {pages} {54--60}
  (\bibinfo {year} {2021})}\BibitemShut {NoStop}%
\bibitem [{\citenamefont {Biamonte}\ \emph {et~al.}(2017)\citenamefont
  {Biamonte}, \citenamefont {Wittek}, \citenamefont {Pancotti}, \citenamefont
  {Rebentrost}, \citenamefont {Wiebe},\ and\ \citenamefont
  {Lloyd}}]{Biamonte2017}%
  \BibitemOpen
  \bibfield  {author} {\bibinfo {author} {\bibfnamefont {J.}~\bibnamefont
  {Biamonte}}, \bibinfo {author} {\bibfnamefont {P.}~\bibnamefont {Wittek}},
  \bibinfo {author} {\bibfnamefont {N.}~\bibnamefont {Pancotti}}, \bibinfo
  {author} {\bibfnamefont {P.}~\bibnamefont {Rebentrost}}, \bibinfo {author}
  {\bibfnamefont {N.}~\bibnamefont {Wiebe}}, \ and\ \bibinfo {author}
  {\bibfnamefont {S.}~\bibnamefont {Lloyd}},\ }\bibfield  {title} {\enquote
  {\bibinfo {title} {Quantum machine learning},}\ }\href {\doibase
  10.1038/nature23474} {\bibfield  {journal} {\bibinfo  {journal} {Nature}\
  }\textbf {\bibinfo {volume} {549}},\ \bibinfo {pages} {195--202} (\bibinfo
  {year} {2017})}\BibitemShut {NoStop}%
\bibitem [{\citenamefont {Kandala}\ \emph {et~al.}(2019)\citenamefont
  {Kandala}, \citenamefont {Temme}, \citenamefont {C{\'o}rcoles}, \citenamefont
  {Mezzacapo}, \citenamefont {Chow},\ and\ \citenamefont
  {Gambetta}}]{Kandala2019}%
  \BibitemOpen
  \bibfield  {author} {\bibinfo {author} {\bibfnamefont {A.}~\bibnamefont
  {Kandala}}, \bibinfo {author} {\bibfnamefont {K.}~\bibnamefont {Temme}},
  \bibinfo {author} {\bibfnamefont {A.~D.}\ \bibnamefont {C{\'o}rcoles}},
  \bibinfo {author} {\bibfnamefont {A.}~\bibnamefont {Mezzacapo}}, \bibinfo
  {author} {\bibfnamefont {J.~M.}\ \bibnamefont {Chow}}, \ and\ \bibinfo
  {author} {\bibfnamefont {J.~M.}\ \bibnamefont {Gambetta}},\ }\bibfield
  {title} {\enquote {\bibinfo {title} {Error mitigation extends the
  computational reach of a noisy quantum processor},}\ }\href {\doibase
  10.1038/s41586-019-1040-7} {\bibfield  {journal} {\bibinfo  {journal}
  {Nature}\ }\textbf {\bibinfo {volume} {567}},\ \bibinfo {pages} {491--495}
  (\bibinfo {year} {2019})}\BibitemShut {NoStop}%
\bibitem [{\citenamefont {Shor}(1997)}]{Shor97}%
  \BibitemOpen
  \bibfield  {author} {\bibinfo {author} {\bibfnamefont {P.~W.}\ \bibnamefont
  {Shor}},\ }\bibfield  {title} {\enquote {\bibinfo {title} {Polynomial-time
  algorithms for prime factorization and discrete logarithms on a quantum
  computer},}\ }\href {\doibase 10.1137/S0097539795293172} {\bibfield
  {journal} {\bibinfo  {journal} {SIAM Journal on Computing}\ }\textbf
  {\bibinfo {volume} {26}},\ \bibinfo {pages} {1484--1509} (\bibinfo {year}
  {1997})},\ \Eprint
  {http://arxiv.org/abs/https://doi.org/10.1137/S0097539795293172}
  {https://doi.org/10.1137/S0097539795293172} \BibitemShut {NoStop}%
\bibitem [{\citenamefont {Dutkiewicz}, \citenamefont {Terhal},\ and\
  \citenamefont {O'Brien}(2022)}]{Dutkiewicz2022}%
  \BibitemOpen
  \bibfield  {author} {\bibinfo {author} {\bibfnamefont {A.}~\bibnamefont
  {Dutkiewicz}}, \bibinfo {author} {\bibfnamefont {B.~M.}\ \bibnamefont
  {Terhal}}, \ and\ \bibinfo {author} {\bibfnamefont {T.~E.}\ \bibnamefont
  {O'Brien}},\ }\bibfield  {title} {\enquote {\bibinfo {title}
  {Heisenberg-limited quantum phase estimation of multiple eigenvalues with few
  control qubits},}\ }\href {\doibase 10.22331/q-2022-10-06-830} {\bibfield
  {journal} {\bibinfo  {journal} {{Quantum}}\ }\textbf {\bibinfo {volume}
  {6}},\ \bibinfo {pages} {830} (\bibinfo {year} {2022})}\BibitemShut {NoStop}%
\bibitem [{\citenamefont {Czarnik}\ \emph {et~al.}(2021)\citenamefont
  {Czarnik}, \citenamefont {Arrasmith}, \citenamefont {Coles},\ and\
  \citenamefont {Cincio}}]{Czarnik2021errormitigation}%
  \BibitemOpen
  \bibfield  {author} {\bibinfo {author} {\bibfnamefont {P.}~\bibnamefont
  {Czarnik}}, \bibinfo {author} {\bibfnamefont {A.}~\bibnamefont {Arrasmith}},
  \bibinfo {author} {\bibfnamefont {P.~J.}\ \bibnamefont {Coles}}, \ and\
  \bibinfo {author} {\bibfnamefont {L.}~\bibnamefont {Cincio}},\ }\bibfield
  {title} {\enquote {\bibinfo {title} {Error mitigation with {C}lifford
  quantum-circuit data},}\ }\href {\doibase 10.22331/q-2021-11-26-592}
  {\bibfield  {journal} {\bibinfo  {journal} {{Quantum}}\ }\textbf {\bibinfo
  {volume} {5}},\ \bibinfo {pages} {592} (\bibinfo {year} {2021})}\BibitemShut
  {NoStop}%
\bibitem [{\citenamefont {Berg}\ \emph {et~al.}(2022)\citenamefont {Berg},
  \citenamefont {Minev}, \citenamefont {Kandala},\ and\ \citenamefont
  {Temme}}]{Berg2022}%
  \BibitemOpen
  \bibfield  {author} {\bibinfo {author} {\bibfnamefont {E.~v.~d.}\
  \bibnamefont {Berg}}, \bibinfo {author} {\bibfnamefont {Z.~K.}\ \bibnamefont
  {Minev}}, \bibinfo {author} {\bibfnamefont {A.}~\bibnamefont {Kandala}}, \
  and\ \bibinfo {author} {\bibfnamefont {K.}~\bibnamefont {Temme}},\ }\href
  {\doibase 10.48550/ARXIV.2201.09866} {\enquote {\bibinfo {title}
  {Probabilistic error cancellation with sparse pauli-lindblad models on noisy
  quantum processors},}\ } (\bibinfo {year} {2022})\BibitemShut {NoStop}%
\bibitem [{\citenamefont {Cai}(2021{\natexlab{b}})}]{Cai2021}%
  \BibitemOpen
  \bibfield  {author} {\bibinfo {author} {\bibfnamefont {Z.}~\bibnamefont
  {Cai}},\ }\bibfield  {title} {\enquote {\bibinfo {title} {Quantum {E}rror
  {M}itigation using {S}ymmetry {E}xpansion},}\ }\href {\doibase
  10.22331/q-2021-09-21-548} {\bibfield  {journal} {\bibinfo  {journal}
  {{Quantum}}\ }\textbf {\bibinfo {volume} {5}},\ \bibinfo {pages} {548}
  (\bibinfo {year} {2021}{\natexlab{b}})}\BibitemShut {NoStop}%
\bibitem [{\citenamefont {LaRose}\ \emph {et~al.}(2022)\citenamefont {LaRose},
  \citenamefont {Mari}, \citenamefont {Kaiser}, \citenamefont {Karalekas},
  \citenamefont {Alves}, \citenamefont {Czarnik}, \citenamefont {El~Mandouh},
  \citenamefont {Gordon}, \citenamefont {Hindy}, \citenamefont {Robertson},
  \citenamefont {Thakre}, \citenamefont {Wahl}, \citenamefont {Samuel},
  \citenamefont {Mistri}, \citenamefont {Tremblay}, \citenamefont {Gardner},
  \citenamefont {Stemen}, \citenamefont {Shammah},\ and\ \citenamefont
  {Zeng}}]{LaRose2022}%
  \BibitemOpen
  \bibfield  {author} {\bibinfo {author} {\bibfnamefont {R.}~\bibnamefont
  {LaRose}}, \bibinfo {author} {\bibfnamefont {A.}~\bibnamefont {Mari}},
  \bibinfo {author} {\bibfnamefont {S.}~\bibnamefont {Kaiser}}, \bibinfo
  {author} {\bibfnamefont {P.~J.}\ \bibnamefont {Karalekas}}, \bibinfo {author}
  {\bibfnamefont {A.~A.}\ \bibnamefont {Alves}}, \bibinfo {author}
  {\bibfnamefont {P.}~\bibnamefont {Czarnik}}, \bibinfo {author} {\bibfnamefont
  {M.}~\bibnamefont {El~Mandouh}}, \bibinfo {author} {\bibfnamefont {M.~H.}\
  \bibnamefont {Gordon}}, \bibinfo {author} {\bibfnamefont {Y.}~\bibnamefont
  {Hindy}}, \bibinfo {author} {\bibfnamefont {A.}~\bibnamefont {Robertson}},
  \bibinfo {author} {\bibfnamefont {P.}~\bibnamefont {Thakre}}, \bibinfo
  {author} {\bibfnamefont {M.}~\bibnamefont {Wahl}}, \bibinfo {author}
  {\bibfnamefont {D.}~\bibnamefont {Samuel}}, \bibinfo {author} {\bibfnamefont
  {R.}~\bibnamefont {Mistri}}, \bibinfo {author} {\bibfnamefont
  {M.}~\bibnamefont {Tremblay}}, \bibinfo {author} {\bibfnamefont
  {N.}~\bibnamefont {Gardner}}, \bibinfo {author} {\bibfnamefont {N.~T.}\
  \bibnamefont {Stemen}}, \bibinfo {author} {\bibfnamefont {N.}~\bibnamefont
  {Shammah}}, \ and\ \bibinfo {author} {\bibfnamefont {W.~J.}\ \bibnamefont
  {Zeng}},\ }\bibfield  {title} {\enquote {\bibinfo {title} {Mitiq: {A}
  software package for error mitigation on noisy quantum computers},}\ }\href
  {\doibase 10.22331/q-2022-08-11-774} {\bibfield  {journal} {\bibinfo
  {journal} {{Quantum}}\ }\textbf {\bibinfo {volume} {6}},\ \bibinfo {pages}
  {774} (\bibinfo {year} {2022})}\BibitemShut {NoStop}%
\bibitem [{\citenamefont {Suchsland}\ \emph {et~al.}(2021)\citenamefont
  {Suchsland}, \citenamefont {Tacchino}, \citenamefont {Fischer}, \citenamefont
  {Neupert}, \citenamefont {Barkoutsos},\ and\ \citenamefont
  {Tavernelli}}]{Suchsland2021}%
  \BibitemOpen
  \bibfield  {author} {\bibinfo {author} {\bibfnamefont {P.}~\bibnamefont
  {Suchsland}}, \bibinfo {author} {\bibfnamefont {F.}~\bibnamefont {Tacchino}},
  \bibinfo {author} {\bibfnamefont {M.~H.}\ \bibnamefont {Fischer}}, \bibinfo
  {author} {\bibfnamefont {T.}~\bibnamefont {Neupert}}, \bibinfo {author}
  {\bibfnamefont {P.~K.}\ \bibnamefont {Barkoutsos}}, \ and\ \bibinfo {author}
  {\bibfnamefont {I.}~\bibnamefont {Tavernelli}},\ }\bibfield  {title}
  {\enquote {\bibinfo {title} {Algorithmic {E}rror {M}itigation {S}cheme for
  {C}urrent {Q}uantum {P}rocessors},}\ }\href {\doibase
  10.22331/q-2021-07-01-492} {\bibfield  {journal} {\bibinfo  {journal}
  {{Quantum}}\ }\textbf {\bibinfo {volume} {5}},\ \bibinfo {pages} {492}
  (\bibinfo {year} {2021})}\BibitemShut {NoStop}%
\bibitem [{\citenamefont {McKay}\ \emph {et~al.}(2018)\citenamefont {McKay},
  \citenamefont {Alexander}, \citenamefont {Bello}, \citenamefont {Biercuk},
  \citenamefont {Bishop}, \citenamefont {Chen}, \citenamefont {Chow},
  \citenamefont {C{\'o}rcoles}, \citenamefont {Egger}, \citenamefont {Filipp}
  \emph {et~al.}}]{mckay2018qiskit}%
  \BibitemOpen
  \bibfield  {author} {\bibinfo {author} {\bibfnamefont {D.~C.}\ \bibnamefont
  {McKay}}, \bibinfo {author} {\bibfnamefont {T.}~\bibnamefont {Alexander}},
  \bibinfo {author} {\bibfnamefont {L.}~\bibnamefont {Bello}}, \bibinfo
  {author} {\bibfnamefont {M.~J.}\ \bibnamefont {Biercuk}}, \bibinfo {author}
  {\bibfnamefont {L.}~\bibnamefont {Bishop}}, \bibinfo {author} {\bibfnamefont
  {J.}~\bibnamefont {Chen}}, \bibinfo {author} {\bibfnamefont {J.~M.}\
  \bibnamefont {Chow}}, \bibinfo {author} {\bibfnamefont {A.~D.}\ \bibnamefont
  {C{\'o}rcoles}}, \bibinfo {author} {\bibfnamefont {D.}~\bibnamefont {Egger}},
  \bibinfo {author} {\bibfnamefont {S.}~\bibnamefont {Filipp}},  \emph
  {et~al.},\ }\bibfield  {title} {\enquote {\bibinfo {title} {Qiskit backend
  specifications for openqasm and openpulse experiments},}\ }\href@noop {}
  {\bibfield  {journal} {\bibinfo  {journal} {arXiv preprint arXiv:1809.03452}\
  } (\bibinfo {year} {2018})}\BibitemShut {NoStop}%
\bibitem [{\citenamefont {Funcke}\ \emph {et~al.}(2022)\citenamefont {Funcke},
  \citenamefont {Hartung}, \citenamefont {Jansen}, \citenamefont {K\"uhn},
  \citenamefont {Stornati},\ and\ \citenamefont {Wang}}]{FunckePRA2022}%
  \BibitemOpen
  \bibfield  {author} {\bibinfo {author} {\bibfnamefont {L.}~\bibnamefont
  {Funcke}}, \bibinfo {author} {\bibfnamefont {T.}~\bibnamefont {Hartung}},
  \bibinfo {author} {\bibfnamefont {K.}~\bibnamefont {Jansen}}, \bibinfo
  {author} {\bibfnamefont {S.}~\bibnamefont {K\"uhn}}, \bibinfo {author}
  {\bibfnamefont {P.}~\bibnamefont {Stornati}}, \ and\ \bibinfo {author}
  {\bibfnamefont {X.}~\bibnamefont {Wang}},\ }\bibfield  {title} {\enquote
  {\bibinfo {title} {Measurement error mitigation in quantum computers through
  classical bit-flip correction},}\ }\href {\doibase
  10.1103/PhysRevA.105.062404} {\bibfield  {journal} {\bibinfo  {journal}
  {Phys. Rev. A}\ }\textbf {\bibinfo {volume} {105}},\ \bibinfo {pages}
  {062404} (\bibinfo {year} {2022})}\BibitemShut {NoStop}%
\bibitem [{\citenamefont {Elben}\ \emph {et~al.}(2020)\citenamefont {Elben},
  \citenamefont {Vermersch}, \citenamefont {van Bijnen}, \citenamefont
  {Kokail}, \citenamefont {Brydges}, \citenamefont {Maier}, \citenamefont
  {Joshi}, \citenamefont {Blatt}, \citenamefont {Roos},\ and\ \citenamefont
  {Zoller}}]{elben2020cross}%
  \BibitemOpen
  \bibfield  {author} {\bibinfo {author} {\bibfnamefont {A.}~\bibnamefont
  {Elben}}, \bibinfo {author} {\bibfnamefont {B.}~\bibnamefont {Vermersch}},
  \bibinfo {author} {\bibfnamefont {R.}~\bibnamefont {van Bijnen}}, \bibinfo
  {author} {\bibfnamefont {C.}~\bibnamefont {Kokail}}, \bibinfo {author}
  {\bibfnamefont {T.}~\bibnamefont {Brydges}}, \bibinfo {author} {\bibfnamefont
  {C.}~\bibnamefont {Maier}}, \bibinfo {author} {\bibfnamefont {M.~K.}\
  \bibnamefont {Joshi}}, \bibinfo {author} {\bibfnamefont {R.}~\bibnamefont
  {Blatt}}, \bibinfo {author} {\bibfnamefont {C.~F.}\ \bibnamefont {Roos}}, \
  and\ \bibinfo {author} {\bibfnamefont {P.}~\bibnamefont {Zoller}},\
  }\bibfield  {title} {\enquote {\bibinfo {title} {Cross-platform verification
  of intermediate scale quantum devices},}\ }\href@noop {} {\bibfield
  {journal} {\bibinfo  {journal} {Physical review letters}\ }\textbf {\bibinfo
  {volume} {124}},\ \bibinfo {pages} {010504} (\bibinfo {year}
  {2020})}\BibitemShut {NoStop}%
\bibitem [{\citenamefont {Lanyon}\ \emph {et~al.}(2017)\citenamefont {Lanyon},
  \citenamefont {Maier}, \citenamefont {Holz{\"a}pfel}, \citenamefont
  {Baumgratz}, \citenamefont {Hempel}, \citenamefont {Jurcevic}, \citenamefont
  {Dhand}, \citenamefont {Buyskikh}, \citenamefont {Daley}, \citenamefont
  {Cramer} \emph {et~al.}}]{lanyon2017efficient}%
  \BibitemOpen
  \bibfield  {author} {\bibinfo {author} {\bibfnamefont {B.}~\bibnamefont
  {Lanyon}}, \bibinfo {author} {\bibfnamefont {C.}~\bibnamefont {Maier}},
  \bibinfo {author} {\bibfnamefont {M.}~\bibnamefont {Holz{\"a}pfel}}, \bibinfo
  {author} {\bibfnamefont {T.}~\bibnamefont {Baumgratz}}, \bibinfo {author}
  {\bibfnamefont {C.}~\bibnamefont {Hempel}}, \bibinfo {author} {\bibfnamefont
  {P.}~\bibnamefont {Jurcevic}}, \bibinfo {author} {\bibfnamefont
  {I.}~\bibnamefont {Dhand}}, \bibinfo {author} {\bibfnamefont
  {A.}~\bibnamefont {Buyskikh}}, \bibinfo {author} {\bibfnamefont
  {A.}~\bibnamefont {Daley}}, \bibinfo {author} {\bibfnamefont
  {M.}~\bibnamefont {Cramer}},  \emph {et~al.},\ }\bibfield  {title} {\enquote
  {\bibinfo {title} {Efficient tomography of a quantum many-body system},}\
  }\href@noop {} {\bibfield  {journal} {\bibinfo  {journal} {Nature Physics}\
  }\textbf {\bibinfo {volume} {13}},\ \bibinfo {pages} {1158--1162} (\bibinfo
  {year} {2017})}\BibitemShut {NoStop}%
\bibitem [{\citenamefont {Flammia}\ and\ \citenamefont
  {Liu}(2011)}]{flammia2011direct}%
  \BibitemOpen
  \bibfield  {author} {\bibinfo {author} {\bibfnamefont {S.~T.}\ \bibnamefont
  {Flammia}}\ and\ \bibinfo {author} {\bibfnamefont {Y.-K.}\ \bibnamefont
  {Liu}},\ }\bibfield  {title} {\enquote {\bibinfo {title} {Direct fidelity
  estimation from few pauli measurements},}\ }\href@noop {} {\bibfield
  {journal} {\bibinfo  {journal} {Physical review letters}\ }\textbf {\bibinfo
  {volume} {106}},\ \bibinfo {pages} {230501} (\bibinfo {year}
  {2011})}\BibitemShut {NoStop}%
\bibitem [{\citenamefont {Zalka}(1999)}]{zalka1999grover}%
  \BibitemOpen
  \bibfield  {author} {\bibinfo {author} {\bibfnamefont {C.}~\bibnamefont
  {Zalka}},\ }\bibfield  {title} {\enquote {\bibinfo {title} {Grover’s
  quantum searching algorithm is optimal},}\ }\href@noop {} {\bibfield
  {journal} {\bibinfo  {journal} {Physical Review A}\ }\textbf {\bibinfo
  {volume} {60}},\ \bibinfo {pages} {2746} (\bibinfo {year}
  {1999})}\BibitemShut {NoStop}%
\bibitem [{\citenamefont {Nishio}\ \emph {et~al.}(2020)\citenamefont {Nishio},
  \citenamefont {Pan}, \citenamefont {Satoh}, \citenamefont {Amano},\ and\
  \citenamefont {Meter}}]{Shin2020}%
  \BibitemOpen
  \bibfield  {author} {\bibinfo {author} {\bibfnamefont {S.}~\bibnamefont
  {Nishio}}, \bibinfo {author} {\bibfnamefont {Y.}~\bibnamefont {Pan}},
  \bibinfo {author} {\bibfnamefont {T.}~\bibnamefont {Satoh}}, \bibinfo
  {author} {\bibfnamefont {H.}~\bibnamefont {Amano}}, \ and\ \bibinfo {author}
  {\bibfnamefont {R.~V.}\ \bibnamefont {Meter}},\ }\bibfield  {title} {\enquote
  {\bibinfo {title} {Extracting success from ibm’s 20-qubit machines using
  error-aware compilation},}\ }\href {\doibase 10.1145/3386162} {\bibfield
  {journal} {\bibinfo  {journal} {J. Emerg. Technol. Comput. Syst.}\ }\textbf
  {\bibinfo {volume} {16}} (\bibinfo {year} {2020}),\
  10.1145/3386162}\BibitemShut {NoStop}%
\bibitem [{\citenamefont {Quetschlich}, \citenamefont {Burgholzer},\ and\
  \citenamefont {Wille}(2023)}]{quetschlich2023}%
  \BibitemOpen
  \bibfield  {author} {\bibinfo {author} {\bibfnamefont {N.}~\bibnamefont
  {Quetschlich}}, \bibinfo {author} {\bibfnamefont {L.}~\bibnamefont
  {Burgholzer}}, \ and\ \bibinfo {author} {\bibfnamefont {R.}~\bibnamefont
  {Wille}},\ }\href@noop {} {\enquote {\bibinfo {title} {Predicting good
  quantum circuit compilation options},}\ } (\bibinfo {year} {2023}),\ \Eprint
  {http://arxiv.org/abs/2210.08027} {arXiv:2210.08027 [quant-ph]} \BibitemShut
  {NoStop}%
\bibitem [{\citenamefont {Proctor}\ \emph {et~al.}(2022)\citenamefont
  {Proctor}, \citenamefont {Rudinger}, \citenamefont {Young}, \citenamefont
  {Nielsen},\ and\ \citenamefont {Blume-Kohout}}]{proctor2022}%
  \BibitemOpen
  \bibfield  {author} {\bibinfo {author} {\bibfnamefont {T.}~\bibnamefont
  {Proctor}}, \bibinfo {author} {\bibfnamefont {K.}~\bibnamefont {Rudinger}},
  \bibinfo {author} {\bibfnamefont {K.}~\bibnamefont {Young}}, \bibinfo
  {author} {\bibfnamefont {E.}~\bibnamefont {Nielsen}}, \ and\ \bibinfo
  {author} {\bibfnamefont {R.}~\bibnamefont {Blume-Kohout}},\ }\bibfield
  {title} {\enquote {\bibinfo {title} {Measuring the capabilities of quantum
  computers},}\ }\href {\doibase 10.1038/s41567-021-01409-7} {\bibfield
  {journal} {\bibinfo  {journal} {Nature Physics}\ }\textbf {\bibinfo {volume}
  {18}},\ \bibinfo {pages} {75--79} (\bibinfo {year} {2022})}\BibitemShut
  {NoStop}%
\bibitem [{\citenamefont {Cross}\ \emph {et~al.}(2019)\citenamefont {Cross},
  \citenamefont {Bishop}, \citenamefont {Sheldon}, \citenamefont {Nation},\
  and\ \citenamefont {Gambetta}}]{cross2019}%
  \BibitemOpen
  \bibfield  {author} {\bibinfo {author} {\bibfnamefont {A.~W.}\ \bibnamefont
  {Cross}}, \bibinfo {author} {\bibfnamefont {L.~S.}\ \bibnamefont {Bishop}},
  \bibinfo {author} {\bibfnamefont {S.}~\bibnamefont {Sheldon}}, \bibinfo
  {author} {\bibfnamefont {P.~D.}\ \bibnamefont {Nation}}, \ and\ \bibinfo
  {author} {\bibfnamefont {J.~M.}\ \bibnamefont {Gambetta}},\ }\bibfield
  {title} {\enquote {\bibinfo {title} {Validating quantum computers using
  randomized model circuits},}\ }\href {\doibase 10.1103/PhysRevA.100.032328}
  {\bibfield  {journal} {\bibinfo  {journal} {Phys. Rev. A}\ }\textbf {\bibinfo
  {volume} {100}},\ \bibinfo {pages} {032328} (\bibinfo {year}
  {2019})}\BibitemShut {NoStop}%
\bibitem [{\citenamefont {Vadali}\ \emph {et~al.}(2023)\citenamefont {Vadali},
  \citenamefont {Kshirsagar}, \citenamefont {Shyamsundar},\ and\ \citenamefont
  {Perdue}}]{vadali2023}%
  \BibitemOpen
  \bibfield  {author} {\bibinfo {author} {\bibfnamefont {A.}~\bibnamefont
  {Vadali}}, \bibinfo {author} {\bibfnamefont {R.}~\bibnamefont {Kshirsagar}},
  \bibinfo {author} {\bibfnamefont {P.}~\bibnamefont {Shyamsundar}}, \ and\
  \bibinfo {author} {\bibfnamefont {G.~N.}\ \bibnamefont {Perdue}},\ }\bibfield
   {title} {\enquote {\bibinfo {title} {Quantum circuit fidelity estimation
  using machine learning},}\ }\href {\doibase 10.1007/s42484-023-00121-4}
  {\bibfield  {journal} {\bibinfo  {journal} {Quantum Machine Intelligence}\
  }\textbf {\bibinfo {volume} {6}},\ \bibinfo {pages} {1} (\bibinfo {year}
  {2023})}\BibitemShut {NoStop}%
\end{thebibliography}%

\end{document}